\documentclass[12pt]{spieman}  % 12pt font required by SPIE;
\usepackage{amsmath,amsfonts,amssymb}
\usepackage{graphicx}
\usepackage{setspace}
\usepackage{tocloft}
\usepackage{amsmath,amsfonts,amssymb}
\usepackage{graphicx}
\usepackage[colorlinks=true, allcolors=blue]{hyperref}
\usepackage{float}

\cftpagenumbersoff{figure}
\cftpagenumbersoff{table} 

\title{Fast, low-noise CCD systems for future strategic X-ray missions}

\author[a,b,*]{Haley R. Stueber}
\author[a]{Tanmoy Chattopadhyay}
\author[a]{Tonya L. Peshel}
\author[a,b]{Abigail Y. Pan}
\author[a,b,c]{Steven W. Allen}
\author[d]{Marshall W. Bautz}
\author[e]{Kevan Donlon}
\author[d]{Catherine E. Grant}
\author[a]{Sven Herrmann}
\author[d]{Beverly J. LaMarr}
\author[d]{Eric D. Miller}
\author[a,c]{R. Glenn Morris}
\author[a]{Peter Orel}
\author[a]{Artem Poliszczuk}
\author[d]{Gregory Y. Prigozhin}
\affil[a]{Kavli Institute for Particle Astrophysics and Cosmology, Stanford University, 452 Lomita Mall, Stanford, CA 94305, USA}
\affil[b]{Department of Physics, Stanford University, 382 Via Pueblo Mall, Stanford CA 94305, USA}
\affil[c]{SLAC National Accelerator Laboratory, 2575 Sand Hill Road, Menlo Park, CA 94025, USA}
\affil[d]{MIT Kavli Institute for Astrophysics and Space Research, Massachusetts Institute of Technology, 70 Vassar St, Cambridge, MA 02139, USA}
\affil[e]{MIT Lincoln Laboratory, 244 Wood St building 1324, Lexington, MA 02421, USA}

\begin{document} 
\maketitle

\begin{abstract}
Future strategic X-ray missions, such as those targeted by the Great Observatories Maturation Program (GOMaP), require fast, low-noise X-ray imaging spectrometers. To achieve the speed and noise capabilities required by such programs, our Stanford team, in collaboration with the MIT Kavli Institute (MKI) and MIT Lincoln Laboratory (MIT-LL), is developing enhanced X-ray charge-coupled devices (CCDs) and readout systems that leverage tailored application-specific integrated circuits (ASICs). Here, we report the energy resolution and noise performance achieved using some of the latest MIT-LL CCDs in conjunction with Stanford-developed Multi-Channel Readout Chip (MCRC) ASICs. Additionally, we present a new sampling method for simultaneous optimization of the output gate (OG), reset gate (RG), and reset drain (RD) biases which, in combination with new integrated fast summing well (SW) and RG clock operation modes, enables the data rates and noise required for future X-ray telescopes. Finally, we present noise power spectral density (PSD) and waveform analysis methods and posit a physical model for characterizing and understanding output stage noise behavior. 
\end{abstract}


{\noindent \footnotesize\textbf{*}H.R. Stueber,  \linkable{hstueber@stanford.edu} }

\begin{spacing}{1} 

\section{INTRODUCTION}
\label{sec:intro}  

Next-generation X-ray observatories such as those targeted by the Great Observatories Maturation Program (GOMaP) will enable transformative studies of the low-luminosity and high-redshift X-ray universe, providing insight into the formation and evolution of black holes, the formation of large scale structure, the roles of gas flows in galaxy formation, and explosive transient phenomena. To accomplish this, these observatories will need to be equipped with sensitive, high spatial resolution, wide-field-of-view imaging detectors. Mitigating pile-up from bright sources and suppressing the impact of the particle background in observations of faint, diffuse sources will be critical, and will require that these imagers achieve high frame rates while maintaining low noise levels and excellent soft energy response. Active-pixel sensors (APS), including hybrid CMOS detectors (HCDs) \cite{HCMOS07, HCMOS17}, monolithic CMOS detectors (MCMOS)\cite{Kenter2018SPIE10762E..09K}, and depleted field-effect transistor (DEPFET) detectors \cite{DEPFET20} such as those developed for the NewAthena Wide Field Imager (WFI) \cite{athenaSPIE2017}, have been shown to deliver excellent performance in some of these areas. However, contemporary HCDs are limited by relatively high readout noise \cite{chattopadhyay18_HCDoverview}, while current DEPFET detectors tend to have relatively large pixel sizes, making them less suitable candidates for missions targeting high spatial resolution over large collecting areas. MCMOS detectors deliver noise performance and speed, but with shallower depletion regions, they lack sensitivity above 6\,keV. Current state-of-the-art X-ray charge-coupled device (CCD) detectors, on the other hand, with their smaller pixels, deeper depletion regions, and low noise levels, have been shown to deliver all of the performance requirements for the next generation of X-ray telescopes, with the exception of frame rates.

The X-ray Astronomy and Observational Cosmology (XOC) group at Stanford, in collaboration with colleagues at the Massachusetts Institute of Technology (MIT) and MIT Lincoln Laboratory (MIT-LL), are working to address this technology gap, developing fast, low-noise X-ray imaging CCDs. In this manuscript, we focus on four distinct approaches being taken to achieve the bandwidth, frame rates, noise, and power requirements of next-generation astronomical X-ray missions:

\begin{itemize}
\item Developing a fast, low-noise CCD output stage to read large-format detectors at high speeds.
\item Engineering application-specific integrated circuit (ASIC) readout electronics that enable parallel readout and provide speed and noise performance comparable to, or better than, discrete electronics solutions at a fraction of their footprint and power consumption. 
\item Optimizing the noise and speed of detector clocks and their driving circuits.
\item Developing efficient, streamlined algorithms to find optimal bias points, and modeling techniques to identify and understand noise sources.
\end{itemize}

\section{FAST, LOW-NOISE X-RAY CCDs}
\label{sec:CCDs}

\begin{figure*}[ht!]
   \begin{center}
   \begin{tabular}{c}
   \includegraphics[width=0.38\textwidth]{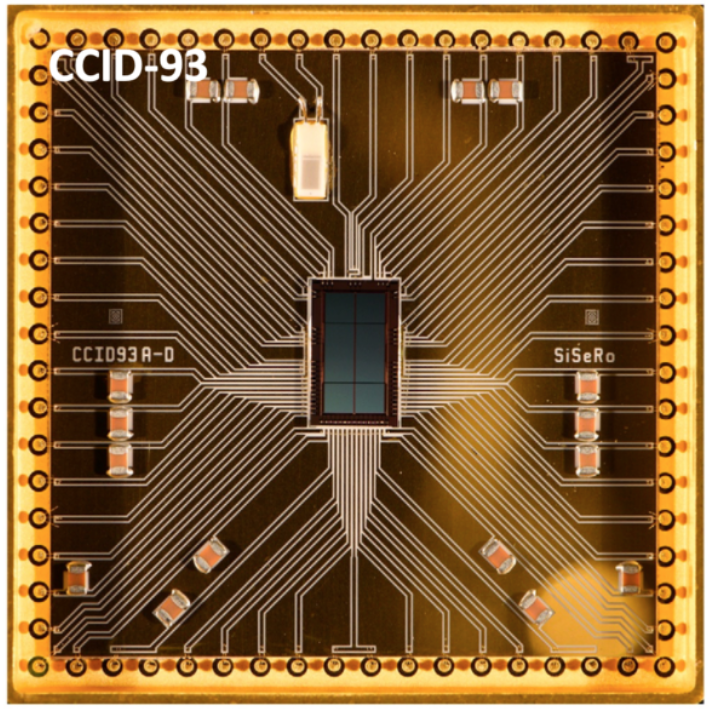}
    \includegraphics[width=0.58\textwidth]{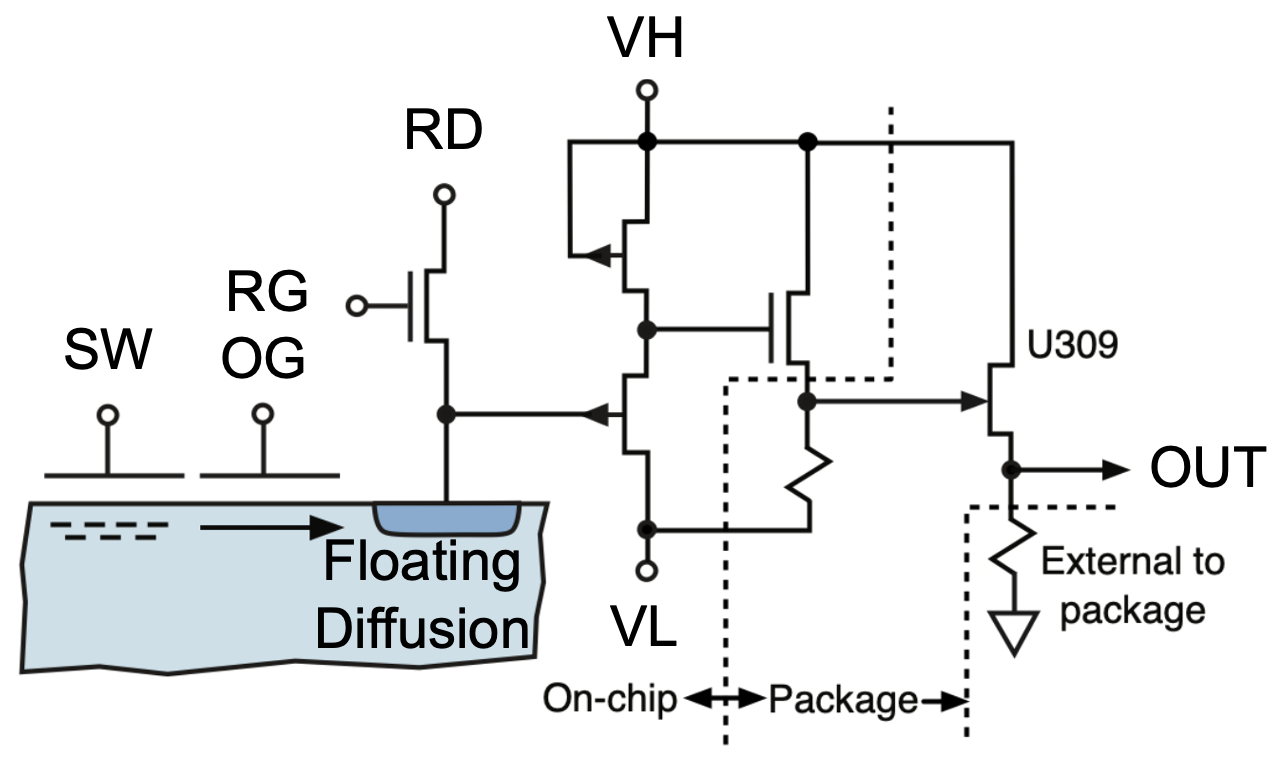}
   \end{tabular}
   \end{center}
   \caption 
   { \label{fig:CCID93_MCRC} 
{\it Left:} Image of an MIT-LL CCID-93 X-ray detector\cite{bautz2024}. The detector has 512x512 8 $\rm \mu m$ pixels in the imaging area, with an identically sized frame store. {\it Right:} JFET output stage of the CCID-93. Charge from the CCD is transferred to the floating diffusion implant, followed by a two-stage output featuring a source follower p-JFET transistor first stage and an n-MOSFET transistor second stage\cite{tanmoyJATIS22}. The reset gate (RG), output gate (OG), and reset drain (RD) nodes are labeled.}
\end{figure*}

The large-area detectors of next-generation observatories will require frame rates at least an order of magnitude faster than those of legacy observatories such as Chandra and XMM-Newton. One of our prototype devices, the MIT-LL CCID-93 detector, is pictured on the left of Figure \ref{fig:CCID93_MCRC}\cite{bautz2024}. Originally intended to demonstrate technology for NASA's Lynx X-ray flagship mission concept\cite{2019lxro.rept.....V}, and later for the Advanced X-ray Imaging Satellite (AXIS)\cite{chrisSPIE2023} probe concept, these detectors address speed and noise requirements that are broadly representative of future large-scale, high-spatial-resolution X-ray observatories.

The MIT-LL CCID-93 is a front-illuminated, small format detector with 8-micron square pixels comprising a 512$\times$512 pixel imaging area and a frame store of the same size. Detailed descriptions of the design and fabrication of the CCID-93 detectors can be found in [\citenum{bautz18}], [\citenum{bautz19}], [\citenum{bautz20}], and [\citenum{gregory2020SPIE}]. In brief, the CCID-93 is a three-phase device in its horizontal and vertical gate structure, where all gates are fabricated with a single layer of polysilicon. Plasma-etched gaps physically isolate the gates. The single polysilicon gate structure allows for faster clock speeds and low clock amplitude swings (as low as 2-3\,V), which translates to lower power consumption and faster readout speeds.

Another critical component is the output stage of the CCID-93. These detectors have two variants of on-chip output amplifier stages, featuring a source follower p-channel junction field-effect transistor (p-JFET) that performs voltage readout, as well as a Single-electron Sensitive Read Out (SiSeRO) output that performs drain current readout. While not the focus of this paper, the SiSeRO architecture and performance results are reported in [\citenum{sisero2021}], [\citenum{sisero2022}], [\citenum{sisero2023}], [\citenum{tanmoyspie2024}], and [\citenum{tanmoySPIE2025}]. The p-JFET output stage is pictured on the right diagram of Figure \ref{fig:CCID93_MCRC}\cite{tanmoyJATIS22}. Charge from the CCD is transferred to the Floating Diffusion implant, which is followed by a two-stage output featuring a fast, high-conversion gain source follower p-JFET first stage and a large-bandwidth n-channel Metal-Oxide-Semiconductor Field-Effect Transistor (n-MOSFET) second stage.

\section{SPEED-READING WITH THE MCRC ASIC}
\label{sec:MCRC}

\begin{figure*}[ht!]
   \begin{center}
   \begin{tabular}{c}
   \includegraphics[width=0.6\textwidth]{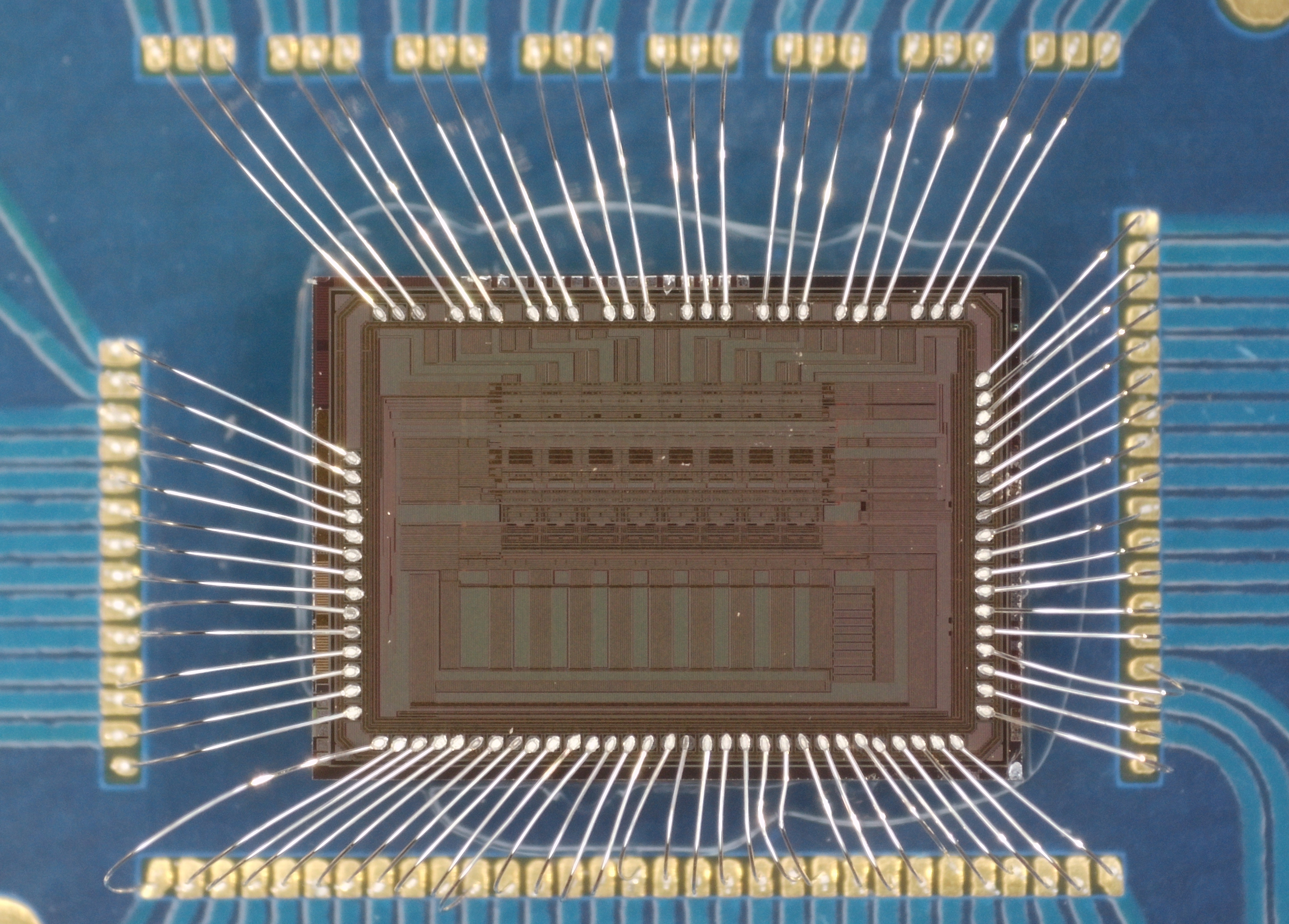}
   \end{tabular}
   \end{center}
   \caption 
   { \label{fig:MCRC} 
The Stanford-developed MCRC-V1 ASIC, featuring 8 channels for fast readout with minimal noise, power consumption and physical footprint\cite{herrmann20_mcrc, porelMCRCspie2022, porelMCRCspie2024}. It has physical dimensions of $\rm 4160\,\mu m$ x $\rm 2900\,\mu m$.
}
\end{figure*}

\noindent To read out these detectors and optimize their speed and noise performances, our team is also developing fast, low-power, low-noise, small-footprint ASIC chips. The Stanford-developed Multi-Channel Readout Chip (MCRC)\cite{herrmann20_mcrc,porelMCRCspie2022} is an analog ASIC used for fast, low-noise readout of MIT-LL CCDs. An image of a fabricated MCRC-V1 chip is pictured in Figure \ref{fig:MCRC}. It has physical dimensions of $\rm 4160\,\mu m \times \rm 2900\,\mu m$ and features 8 analog readout channels that operate in parallel. Full details of the architecture and fabrication of the MCRC ASIC can be found in [\citenum{herrmann20_mcrc}] and [\citenum{porelMCRCspie2022}] with the latest performance reported in [\citenum{porelMCRCspie2024}]. Each of the analog channels of the MCRC consists of an input stage that provides a suitable interface (biasing and impedance) for the output stage of the detector. The user can select between voltage or current inputs for either p-JFET- or SiSeRO-based detector outputs. The input stage is followed by a preamplifier that converts signals from single-ended to differential and provides user-selectable gain settings of 8 or 16 V/V, respectively. The preamplifier output is buffered by a unit-gain, fully differential output buffer, designed to drive a 100 $\Omega$ transmission line up to 1 meter in length. The MCRC signal waveform is sampled by an analog-to-digital converter (ADC), and image data is extracted via a digital pulse processing (DPP) algorithm. Both the ADC and DPP are hosted by an Archon controller\footnote{http://www.sta-inc.net/archon/}. A digital serial peripheral interface (SPI) is used to program the ASIC settings, including control of the internal switch logic and the digital-to-analog converters (DACs) that provide biasing to the internal analog circuitry. 

Integrating multiple functions, the MCRC ASIC streamlines CCD board design by reducing the physical footprint and number of discrete components. It has a large bandwidth and matches the speed and noise performance of our best discrete readout solutions for a fraction of the power consumption\cite{porelMCRCspie2022, porelMCRCspie2024}. The unique features of the ASIC give it the capability to deliver and exceed the baseline requirements in both speed and noise for future large-format X-ray imagers, while enabling fast, parallel readout of multiple outputs.

\section{Speed and Noise Optimization with Onboard Clocks}
\label{sec:SWRG}

\begin{figure*}[ht!]
   \begin{center}
   \begin{tabular}{c}
   \includegraphics[width=0.95\textwidth]{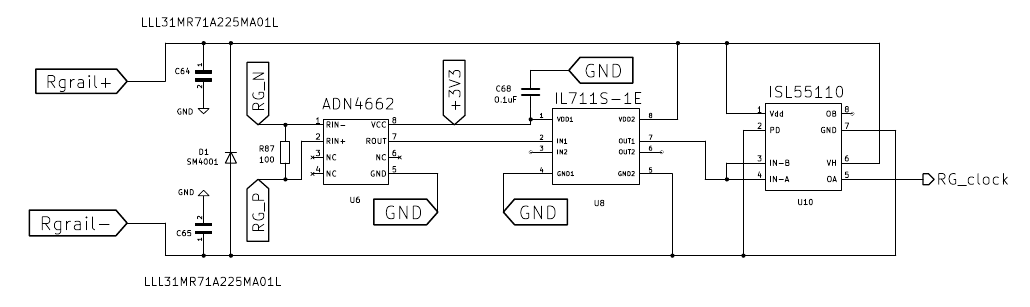}
   \end{tabular}
   \end{center}
   \caption 
   { \label{fig:RG_circuit} 
Circuit schematic for the integrated fast Reset Gate (RG) clock driver. Differential input signals are converted to single-ended, and the negative input is referenced to ground.}
\end{figure*}

\noindent The Reset Gate (RG) and Summing Well (SW) clocks provided by the external Archon controller initially limited the readout speed of our prototype CCID-93 test devices to 4\,MPixel/s\cite{tanmoyJATIS22}. To enable faster serial transfer speeds and reduce noise levels, the SW and RG clock signals have been transitioned to a locally buffered scheme. The onboard circuit for the RG clock signal is pictured in Figure \ref{fig:RG_circuit}; an identical circuit is used to source the SW clock. The onboard circuit converts the clock signals from differential to single-ended form. The single-ended clocks are opto-isolated to enable ground shifting and buffered to provide adjustable amplitudes and offsets for ideal point-of-load driving of the CCD clock inputs.

The high-speed and high-current driving capabilities of the new onboard clock circuitry, combined with a lack of termination, initially introduced ringing in the clock signals. This was mitigated with a ``snubber" circuit. A snubber is a form of termination, which in this case is composed of an RC circuit that attenuates the reflected signal and dampens the ringing without significantly sacrificing the clock speed. The RC values were determined based on the measured parasitic parallel capacitance and series inductance, which was derived from the oscillations measured in the clocks. 

\begin{figure*}[ht!]
   \begin{center}
   \begin{tabular}{c}
   \includegraphics[width=0.8\textwidth]{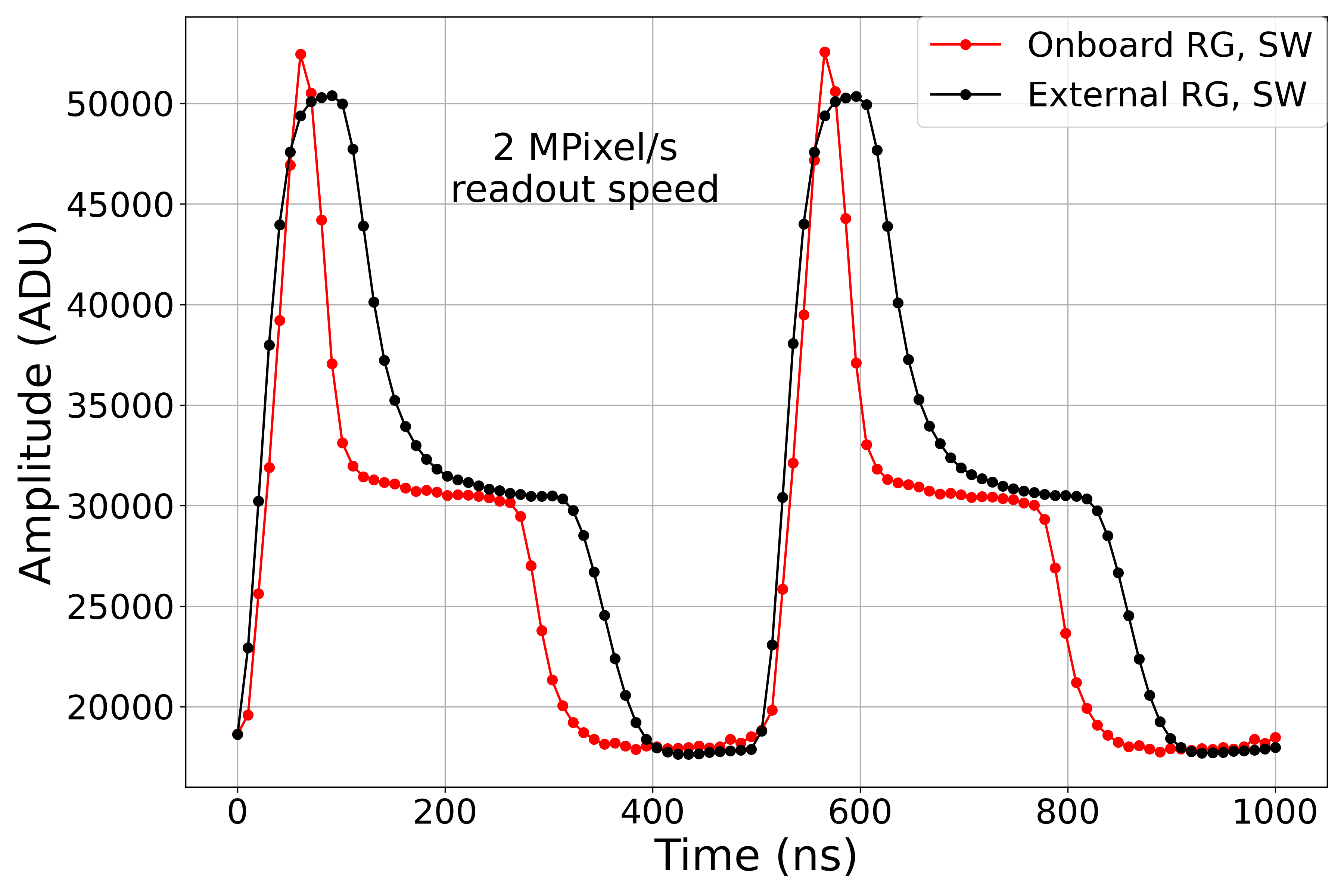}
   \end{tabular}
   \end{center}
   \caption 
   { \label{fig:WF} 
2 MPixel/s waveforms using the external RG and SW drivers (black) and the onboard RG and SW drivers (red). We see $\sim20\%$ more samples in the baseline and the signal region of the waveform using the fast onboard drivers.}
\end{figure*}

A comparison of the CCD waveforms obtained using the externally driven SW and RG clocks (black) and the optimized onboard SW and RG clocks (red) is shown in Figure \ref{fig:WF}. We see a $\rm \sim\!20\%$ recovery of the baseline and signal samples using the faster onboard clocks. These additional samples enable faster serial transfer speeds of up to 5\,MPixel/s. In addition to enhancing the frame rate capabilities of the detectors, the onboard clocks also improve noise performance at lower speeds. The close proximity of the onboard drivers to the output circuitry reduces electromagnetic interference, improving the noise from the externally driven solution. 

\section{Noise Optimization with OG, RG, RD Bias Scanning}
\label{sec:param_scan}

\noindent Finding an optimal bias point for CCD detectors can be challenging, as the relationship between the various bias parameters and detector noise and gain can be complex. The performance of the detector is sensitive to the reset gate high- and low-state voltages (RGH and RGL, respectively), the output gate (OG) voltage, and the reset drain (RD) voltage. The locations of each of these bias nodes in the output stage of the CCID-93 detectors are indicated in the right diagram of Figure \ref{fig:CCID93_MCRC}. The optimal combination of biases can also vary for different detectors, different test systems and readout electronics, and different temperatures. We describe here an approach for performing scans of a grid of values for RGH, RGL, OG, and RD that can be used to efficiently find an optimal operating point. 

\begin{figure*}[ht!]
   \begin{center}
   \begin{tabular}{c}
   \includegraphics[width=0.65\textwidth]{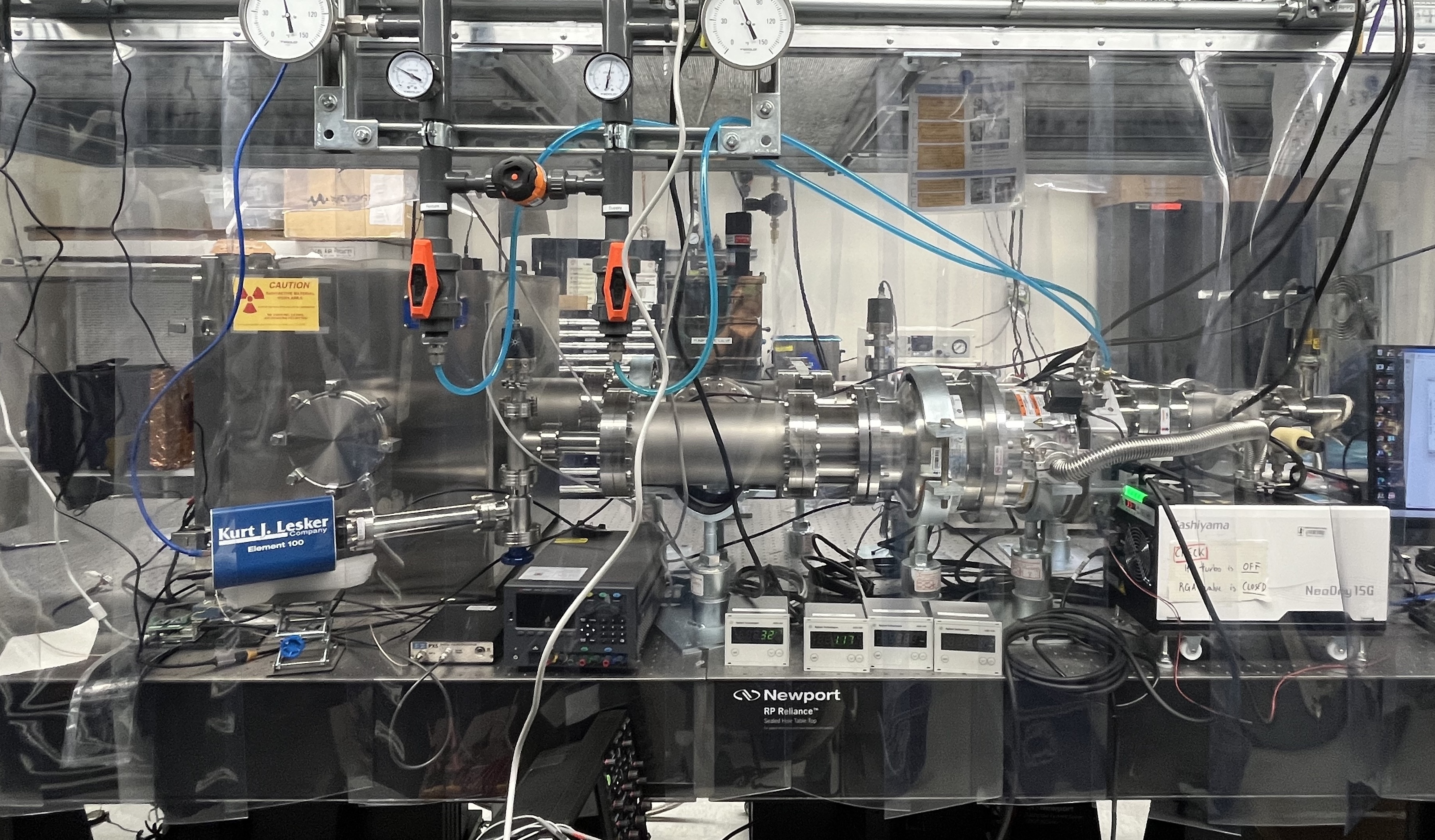}
   \includegraphics[width=0.29\textwidth]{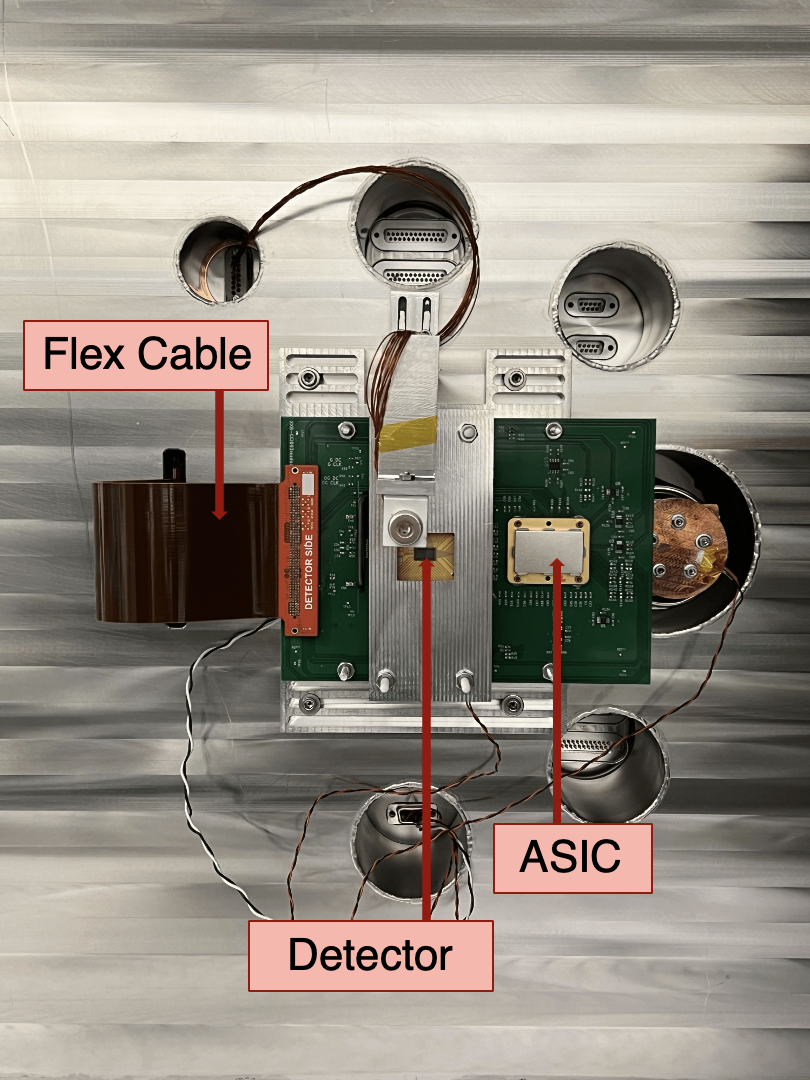}
   \end{tabular}
   \end{center}
   \caption 
   { \label{fig:beamline} 
{\it Left:} Side view of the Gen 2.0 XOC X-ray Beamline CCD test chamber. {\it Right:} CCID-93 detector mounted inside the beamline with the MCRC ASIC and vacuum potted flex cable for transfer of signals between the detector (in vacuum) and the Archon (in atmosphere). Details of the experimental setup can be found in [\citenum{2025SPIEAbbyBeamline}].
}
\end{figure*}

Our scanning algorithm takes start, stop, and step voltage values for each of the bias parameters as inputs. It steps through each combination of RGH, RGL, OG, and RD in the four-dimensional grid, sending each bias configuration to the Archon to set the corresponding voltages, and takes five frames of data at each point. To estimate the read noise, we compute the standard deviation in analog-to-digital units (ADUs) of the 50-column overscan region in each frame, then take the average across the five frames. Because the overscan consists of overclocked pixels with negligible thermal leakage, the measured variation reflects only the system's read noise. Assuming that the relationship between ADUs and electron noise is relatively well-behaved over the range of parameters under test, we find an optimum bias point corresponding to the minimum read noise value in ADUs. We filter out low noise outliers, as they reflect a collapse in the gain, which is otherwise relatively stable over the range of parameters used.
Each scan measures read noise values for around 1200 combinations of parameter values, taking a total of around 2.5 hours to complete. 
We perform the scans in the Gen 2.0 XOC X-ray Beamline vacuum test chamber (Fig. \ref{fig:beamline}, see [\citenum{2025SPIEAbbyBeamline}] for details of the beamline test system) at temperatures ranging from 173\,K to 273\,K in increments of 10\,K. In Section \ref{sec:results}, we present the resulting optimal bias parameters and corresponding read noises and full width half maxima (FWHM) of the single-pixel event spectra Fe-55 $\rm K\alpha$ 5.9\,keV line, and compare this to the read noises and FWHM obtained with a default set of bias parameters.

The purpose of this procedure is to quickly and straightforwardly converge on an ideal operating point for any given detector. Ultimately, this scanning algorithm could be used to efficiently optimize the bias for the large-format, multi-channel detectors required of future X-ray imaging satellites.

\section{Results}
\label{sec:results}

\begin{figure*}[ht!]
\centering
    \includegraphics[width=0.49\linewidth]{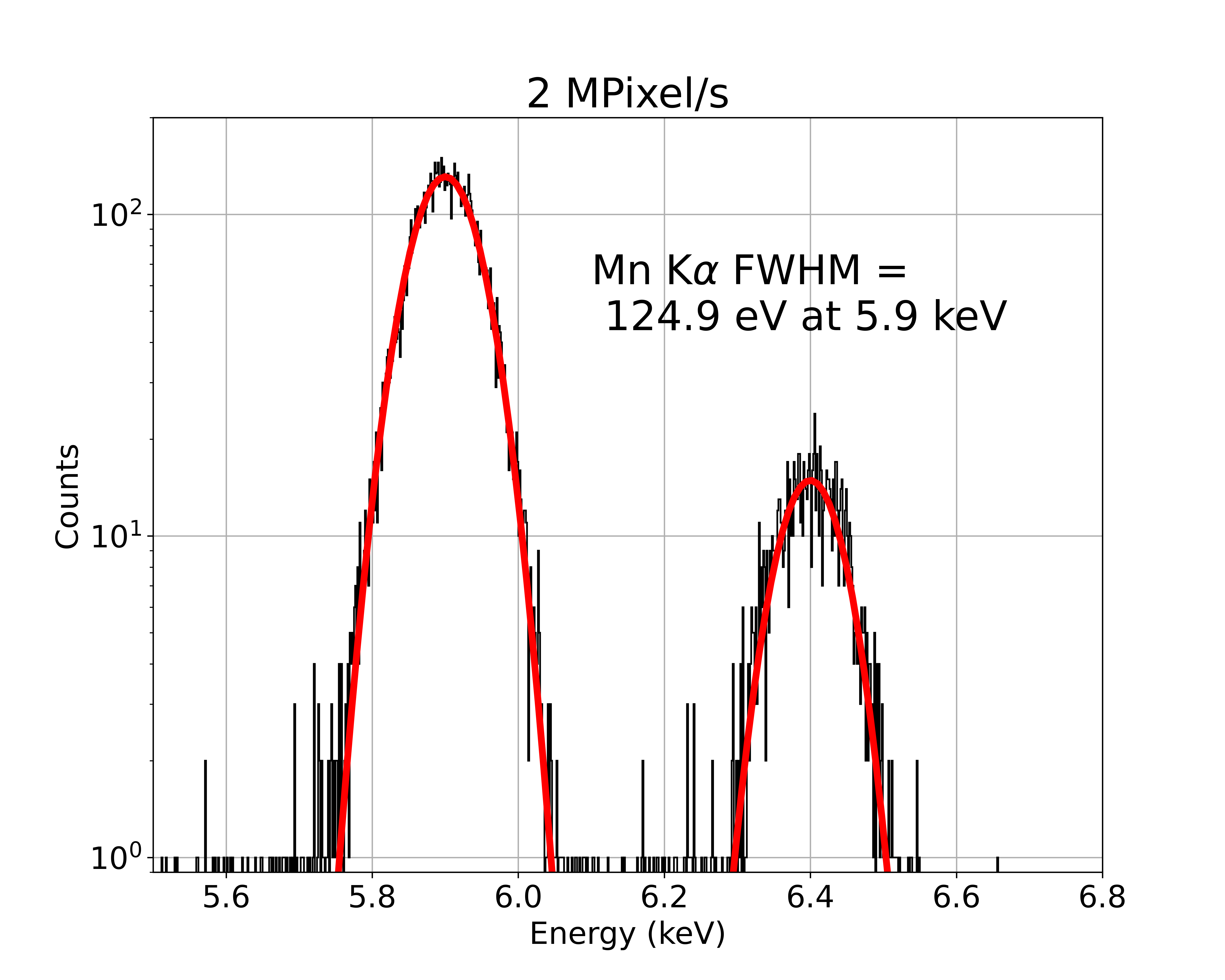}
    \includegraphics[width=0.49\linewidth]{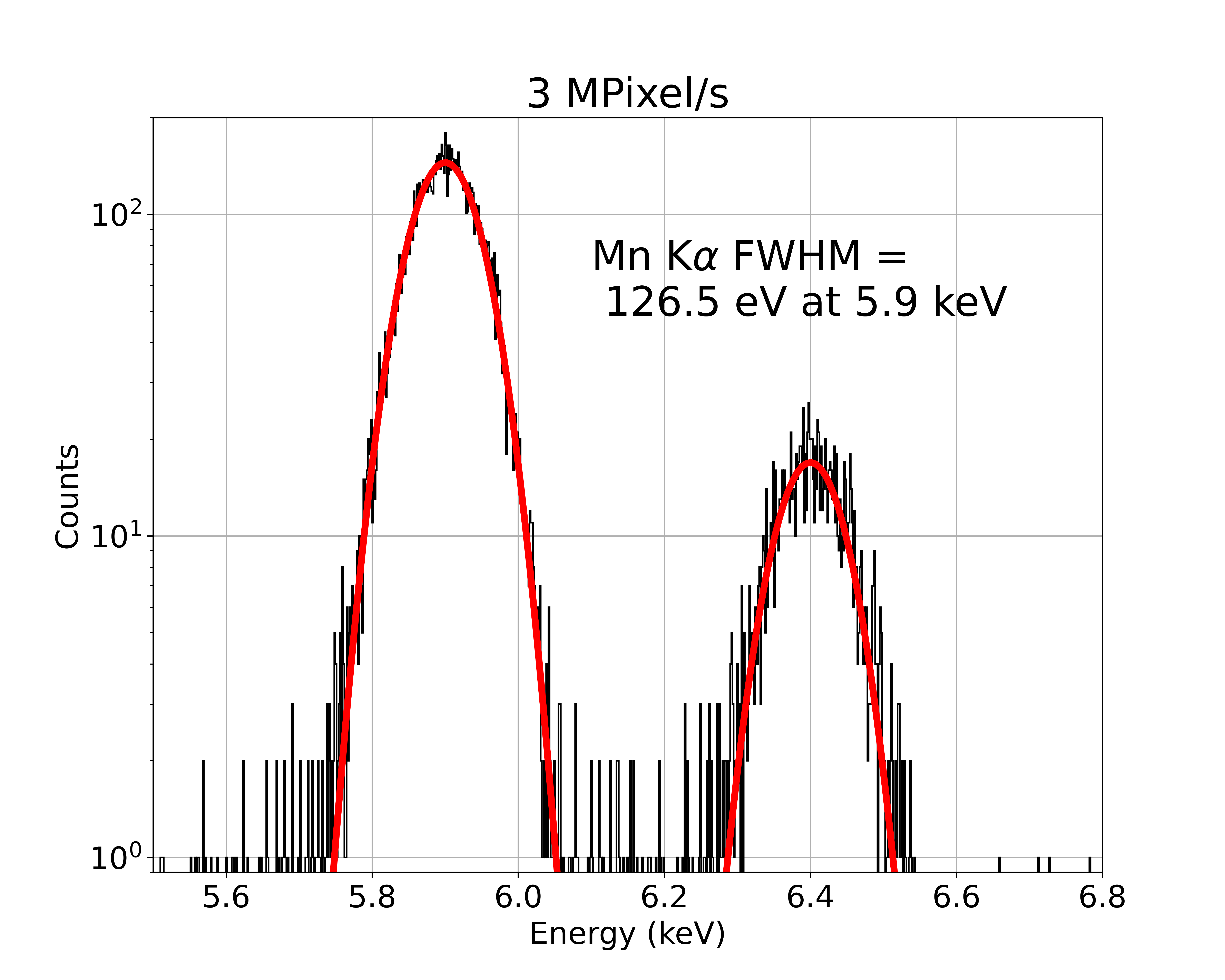}
    \includegraphics[width=0.49\linewidth]{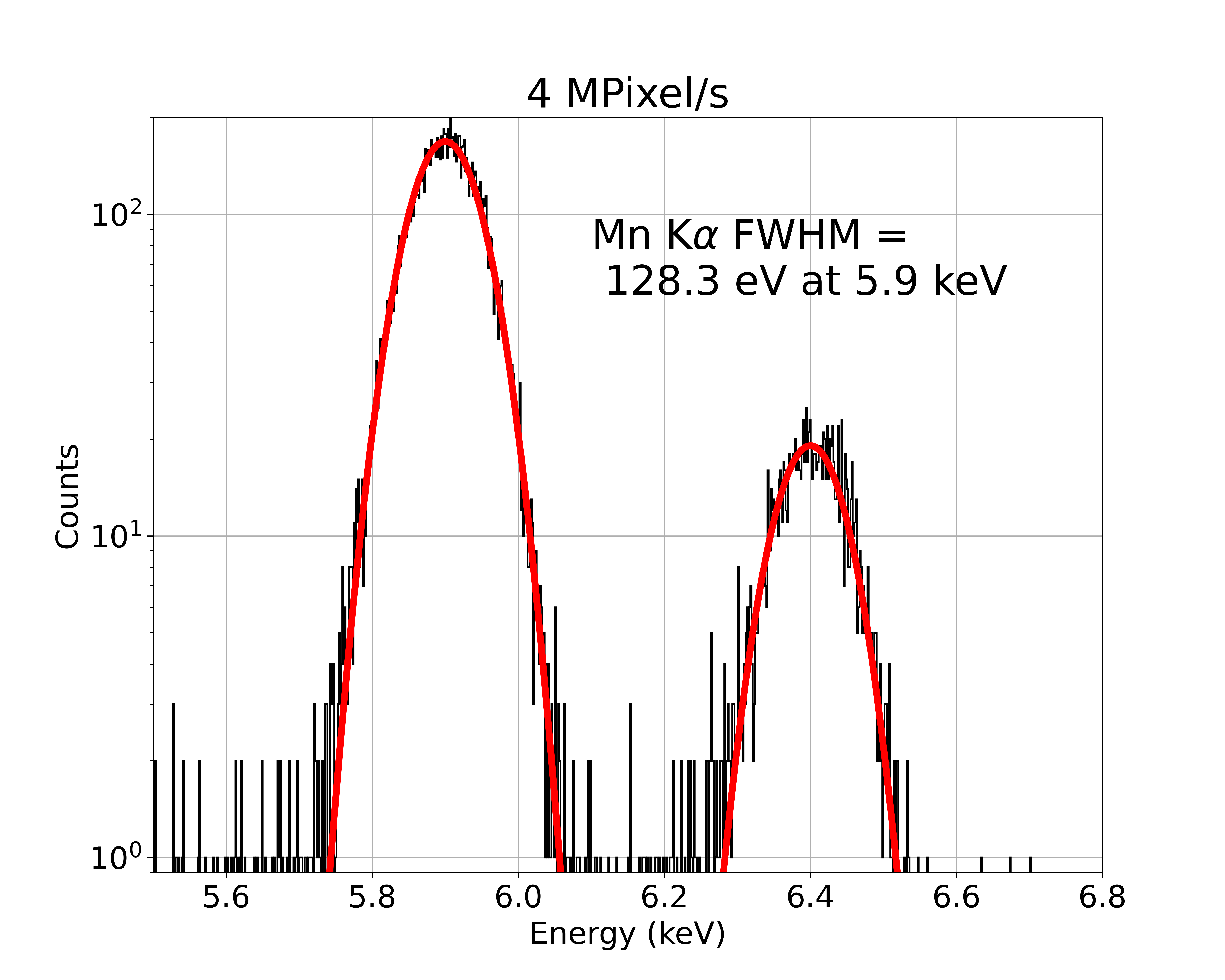}
    \includegraphics[width=0.49\linewidth]{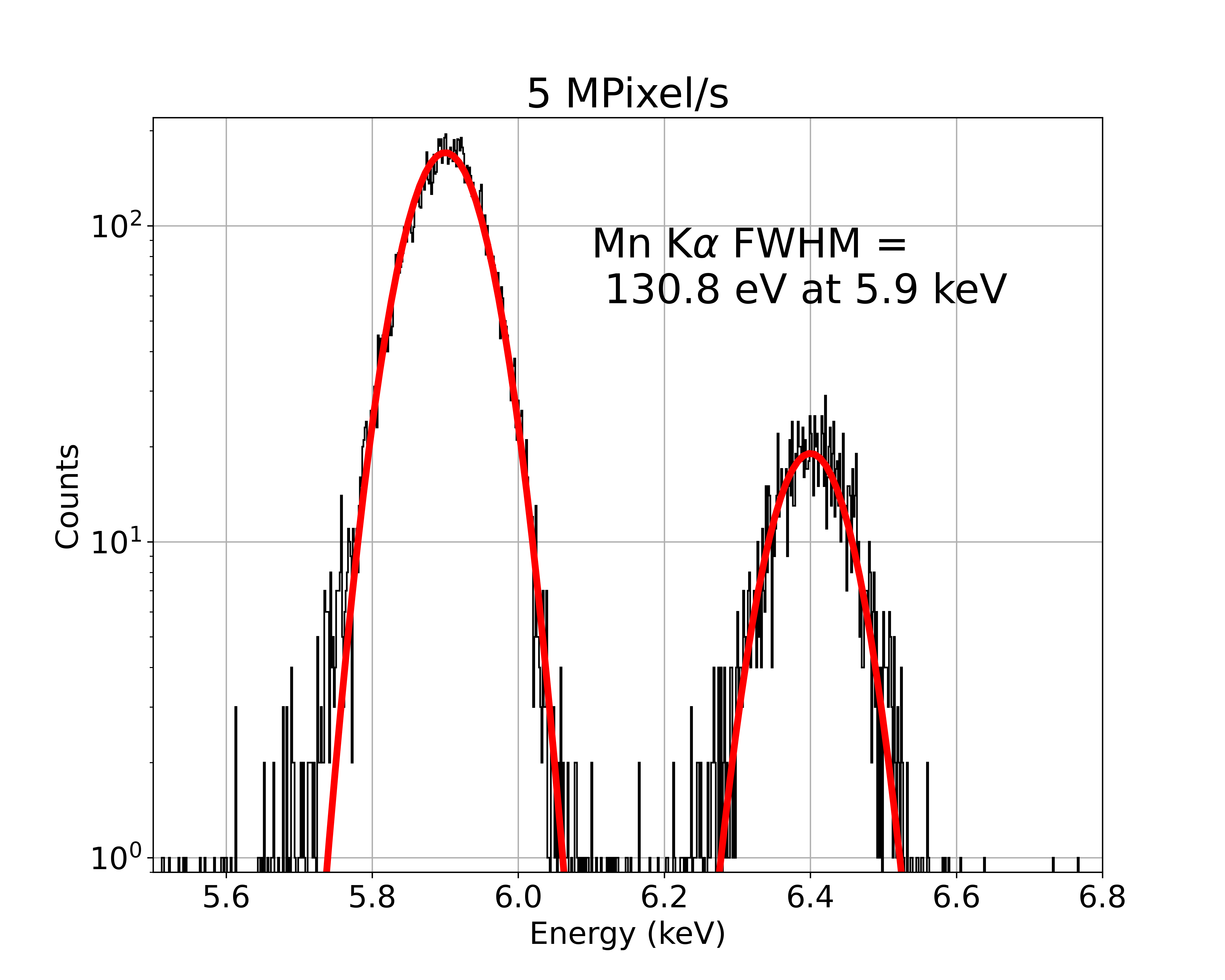}
\caption
{\label{fig:-100_spectrum}
Single-pixel event spectra obtained for an Fe-55 radioactive source at 173\,K from the CCID-93 detector with ASIC readout at 2 Mpixel/s (upper left), 3\,Mpixel/s (upper right), 4\,Mpixel/s (lower left), and 5\,Mpixel/s (lower right) serial transfer speeds.
}
\end{figure*}

\noindent Combining improvements to clock speeds from the implementation of onboard SW and RG clocks outlined in Section \ref{sec:SWRG} with the parameter scanning optimization methods described in Section \ref{sec:param_scan}, Figure \ref{fig:-100_spectrum} presents the optimized single-pixel event spectra obtained from an Fe-55 radioactive source using an MIT-LL CCID-93 and MCRC ASIC readout for serial transfer rates of 2, 3, 4, and 5\,Mpixel/s. For these results, the detector was cooled to 173\,K in the Gen 2.0 XOC X-ray Beamline.

\begin{figure*}[ht!]
   \begin{center}
   \begin{tabular}{c}
   \includegraphics[width=0.8\textwidth]{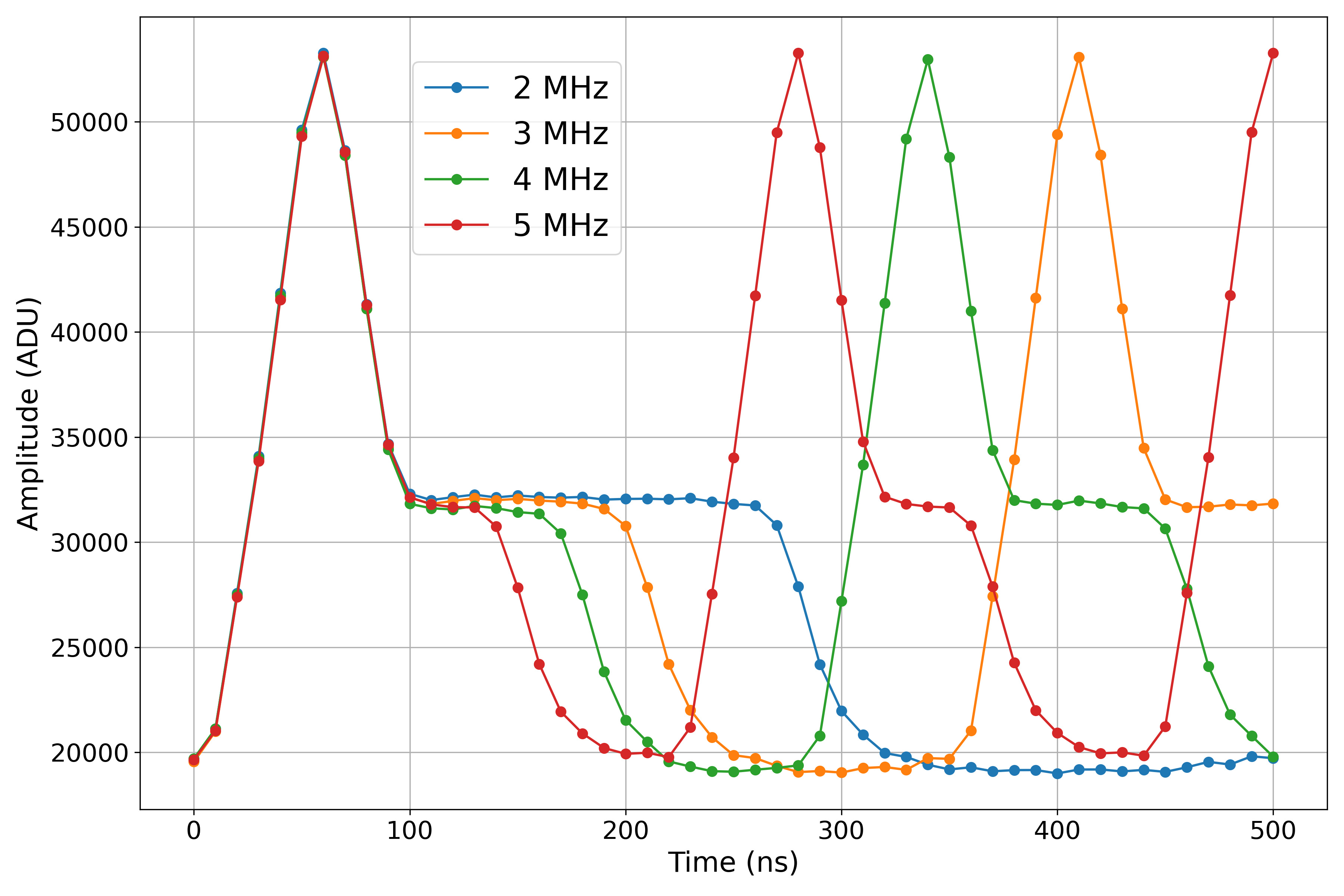}
   \end{tabular}
   \end{center}
   \caption
   { \label{fig:2MHz_5MHz_WF} 
Comparison of 2 MPixel/s, 3 MPixel/s, 4 MPixel/s, and 5 MPixel/s readout rate waveforms from the CCID-93 detector. Fast readout rates have been enabled by our use of onboard fast SW and RG clocks.  
}
\end{figure*}

A summary of the read noises and Fe-55 $\rm K\alpha$ (5.9\,keV) single-pixel event spectra FWHMs in eV for the different serial transfer speeds is given in Table \ref{tab:SWRG_noise}, while Figure \ref{fig:2MHz_5MHz_WF} presents the corresponding raw waveforms. 

\begin{table}[H]
    \caption{Best read noise achieved with the CCID-93 detector and ASIC readout at serial transfer speeds ranging from 2 Mpixel/s to 5 MPixel/s at 173\,K using the onboard RG and SW clocks described in Section \ref{sec:SWRG}.}
    \label{tab:SWRG_noise}
    \begin{center}
    \begin{tabular}{|c|c|c|}
    \hline
    \textbf{Serial Speed (MPixel/s)} & {\textbf{Noise (e$^-$)}} & {\textbf{FWHM (eV) at 5.9\,keV}} \\
    \hline
    2 & $2.18\pm0.01$ & $124.9\pm1.0$ \\
    3 & $2.67\pm0.02$ & $126.5\pm0.9$ \\
    4 & $3.07\pm0.01$ & $128.3\pm0.9$ \\
    5 & $3.85\pm0.02$ & $130.8\pm0.9$ \\
    \hline
    \end{tabular}
    \end{center}
\end{table}

\begin{figure*}[ht!]
   \begin{center}
   \begin{tabular}{c}
   \includegraphics[width=0.95\textwidth]{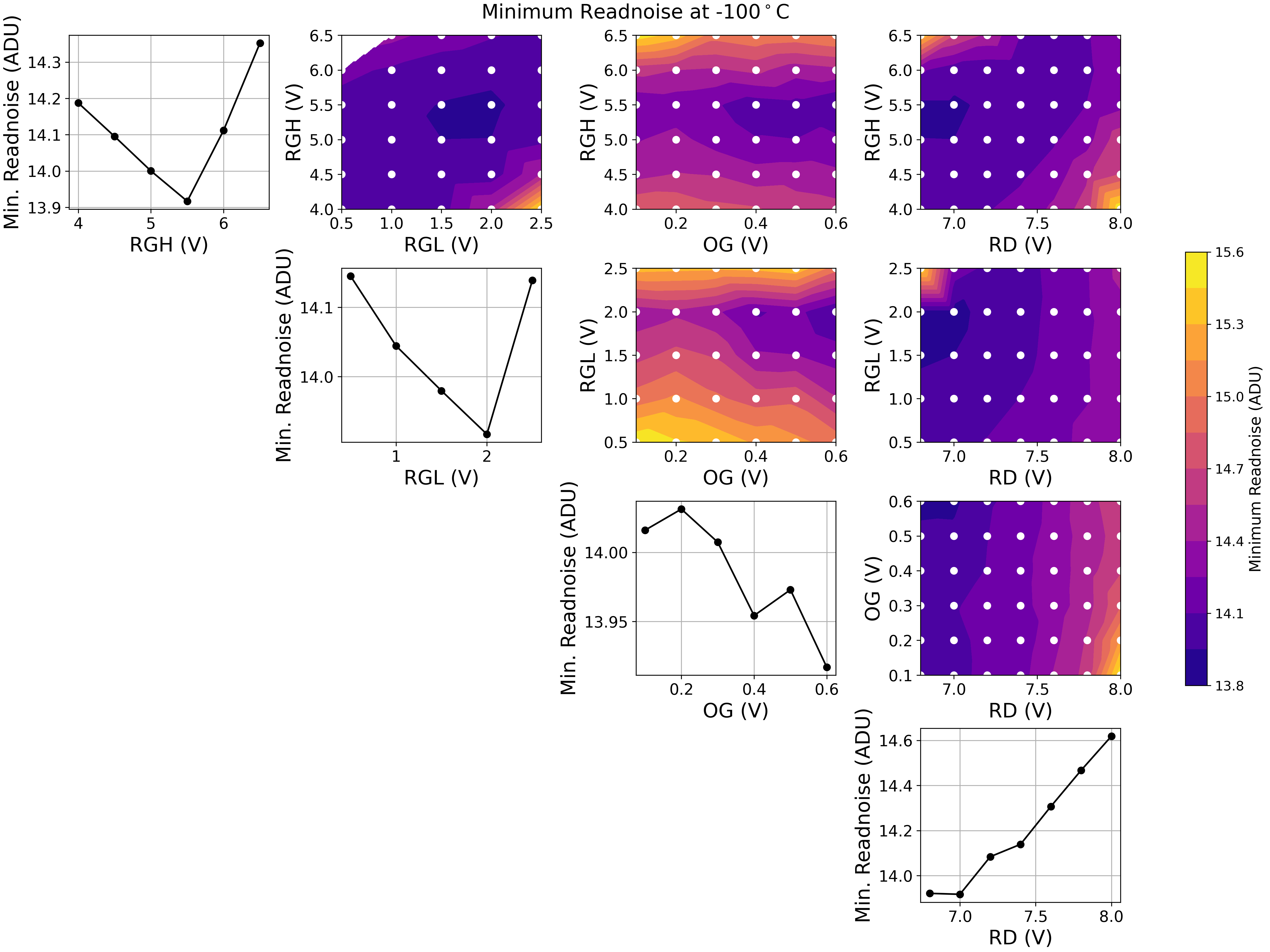}
   \end{tabular}
   \end{center}
   \caption 
   { \label{fig:-100_triangle} 
Summary plots from a scan over RGH, RGL, OG, and RD. In the diagonal plots, the minimum read noise (in ADU) is reported at each value of RGH, RGL, OG, and RD that was scanned over. The off-diagonal plot elements show two-dimensional, linearly interpolated contour plots of the minimum read noise recorded for each combination of the four scan parameters. The white points on each contour plot represent parameter values where read noise values were actually measured. Gaps in the contours, particularly for RGH=6.5\,V and RGL=0.5\,V, indicate combinations of parameter values that were not allowed by the scan and for which no read noise values were recorded.
}
\end{figure*}

Figure \ref{fig:-100_triangle} presents triangle plots of the optimal read noises in the ranges of values given for the RGH, RGL, OG, and RD, parameter scan at 173\,K. In each diagonal plot, the recorded values are the minima at the given bias parameter for any combination of the other three parameters. The off-diagonal, two-dimensional, linearly interpolated contour plots trace the minimum read noise over a grid of any combination of two given parameters, in order to visualize whether there may be a simply varying relationship or degeneracy between two sets of parameters, or whether a local or global minimum in the read noise has been found in that space. Gaps in the contours indicate combinations of parameter values that were not allowed by the scan, and for which no corresponding read noise was recorded. 

A summary of the optimal RGH, RGL, OG, and RD bias parameters found using the scanning procedure defined in Section \ref{sec:param_scan} can be found in Table \ref{tab:bias_params}. We find that the optimal bias voltages for RG, OG, and RD vary with temperature. While different combinations of OG and RD could yield similar noise performance, the optimal RGH and RGL values were more sensitive to the temperature of the detector. RGH appears to prefer increasingly positive voltages as the temperature increases from 173\,K to 243\,K, and then lower voltages from 243\,K to 273\,K. RGL seems to prefer lower voltages at higher temperatures, with the exception of 243\,K. The results suggest that optimizing the bias of these detectors at their intended operating temperatures will be essential in achieving the best possible performance with future X-ray observatories.

\begin{figure*}[ht!]
   \begin{center}
   \begin{tabular}{c}
   \includegraphics[width=0.95\textwidth]{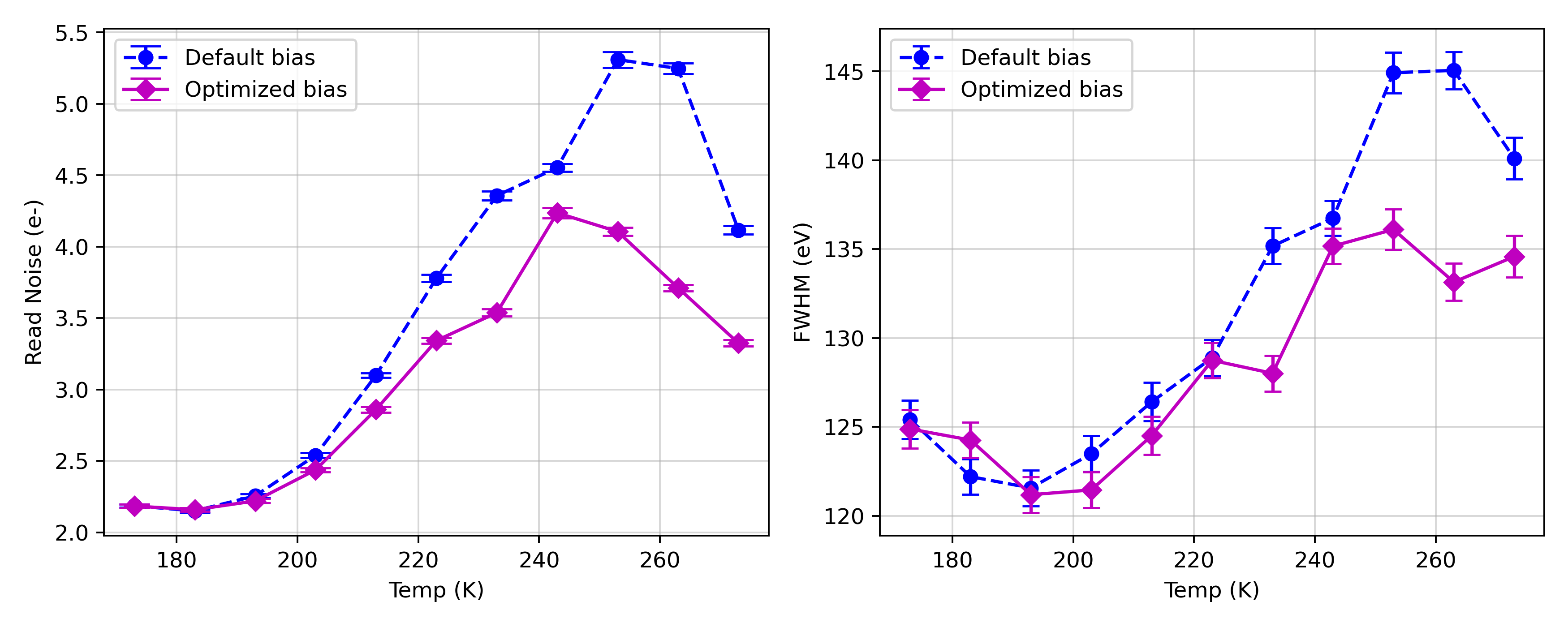}
   \end{tabular}
   \end{center}
   \caption 
   { \label{fig:scan_temp_noise_fwhm} 
{\it Left:} Plot of the read noise in electrons as a function of temperature, for the default bias parameters of RGH=5.0\,V, RGL=2.0\,V, OG=0.4\,V, RD=7.3\,V (blue) and for the optimized bias parameters listed in Table 2 (magenta). {\it Right:} Plot of the FWHM of the single-pixel event Fe-55 $\rm K\alpha$ line as a function of temperature, for the default bias parameters (blue) and for the optimized bias parameters (magenta).
}
\end{figure*}

The plot on the left of Figure \ref{fig:scan_temp_noise_fwhm} presents the read noise in electrons for the default set of parameters (RGH=5.0\,V, RGL=2.0\,V, OG=0.4\,V, and RD=\,7.3V) in blue, as well as the optimized set of parameters found at each temperature in magenta. The plot on the right presents the FWHM in eV of the single-pixel event 5.9\,keV Fe-55 $\rm K\alpha$ line as a function of temperature using the same color scheme.

\begin{table}[ht!]
    \caption{Optimized bias settings (RGH, RGL, OG, RD) found at different temperatures for the CCID-93 detector using the methods described in Section \ref{sec:param_scan}.}
    \label{tab:bias_params}
    \begin{center}
    \begin{tabular}{|c|c|c|c|c|}
    \hline
    \textbf{Temp (K)} & \textbf{RGH (V)} & \textbf{RGL (V)} & \textbf{OG (V)} & \textbf{RD (V)} \\
    \hline
    273 & 4.5 & 0.5 & 0.2 & 7.4 \\
    263 & 6.0 & 0.5 & 0.2 & 8.0 \\
    253 & 6.0 & 0.5 & 0.2 & 8.0 \\
    243 & 6.5 & 2.5 & 0.5 & 7.6 \\
    233 & 6.5 & 1.5 & 0.4 & 8.0 \\
    223 & 6.0 & 2.0 & 0.6 & 7.0 \\
    213 & 5.5 & 2.0 & 0.6 & 6.8 \\
    203 & 5.5 & 2.0 & 0.6 & 6.8 \\
    193 & 5.5 & 2.0 & 0.6 & 6.8 \\
    183 & 5.5 & 2.0 & 0.5 & 6.8 \\
    173 & 5.5 & 2.0 & 0.6 & 7.0 \\
    \hline
    \end{tabular}
    \end{center}
\end{table}

\section{Discussion}
\label{sec:discussion}

\noindent The spectral and noise performance of the CCID-93 detectors show substantial promise towards meeting the requirements for next-generation X-ray imaging telescopes. The two-stage output and ASIC readout architectures combined with fast clocking and bias optimization schemes presented here enable readout speeds up to 5\,MPixel/s while maintaining less than $4\,e^-$ of read noise. However, we note some interesting features in the noise behavior with temperature. In particular, the read noise vs. temperature trend for both the optimized and default biases (Fig. \ref{fig:scan_temp_noise_fwhm}) deviates from an expected power law trend with temperature, appearing to show a resonance feature at around 240--260\,K.  

In this section, we analyze raw CCD waveforms and output stage noise PSDs to identify and characterize the sources contributing to the observed trend in read noise with temperature. We then present a physical model based on the temperature and energy dependence of output stage charge carrier trapping/detrapping as a potential explanation for the noise resonance we see at higher temperatures.

\subsection{Raw Waveform and Output Stage PSD Analysis}
\label{sec:noise_temp_discussion}

\begin{figure*}[!t]
   \begin{center}
   \begin{tabular}{c}
   \includegraphics[width=0.95\textwidth]{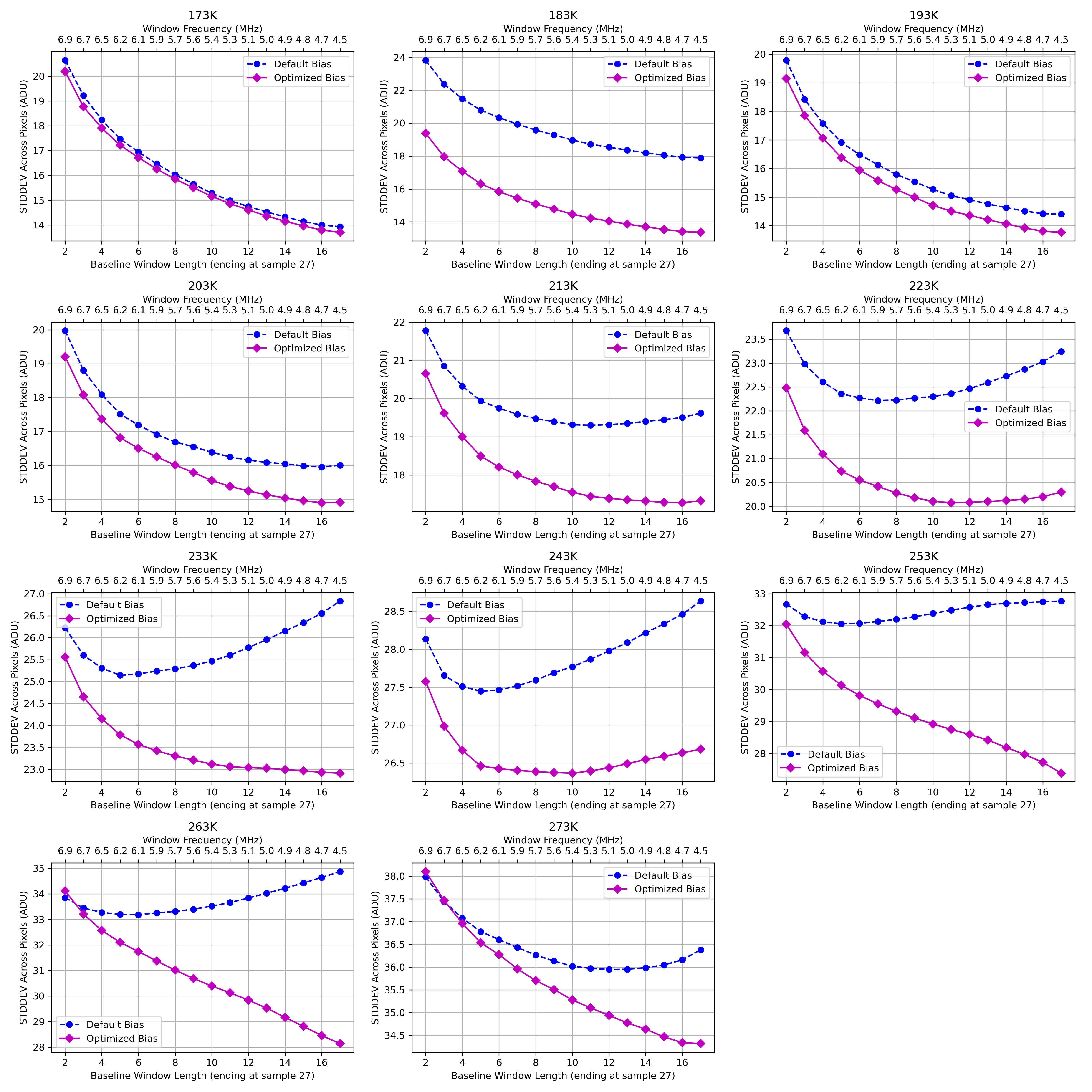}
   \end{tabular}
   \end{center}
   \caption 
   { \label{fig:expanding_baseline} 
Plots of the standard deviation of the baseline-signal waveform amplitude versus number of baseline samples. The last baseline sample is fixed at sample 27 (see Fig. \ref{fig:2MHz_5MHz_WF}) and the first baseline sample is decremented by one until reaching sample 11. Each plot presents the standard deviations for the waveforms obtained using the default bias (blue) and using the optimized bias (magenta) for temperatures between 173\,K and 273\,K.
}
\end{figure*}

\noindent Figure \ref{fig:expanding_baseline} plots the standard deviation of the baseline-signal amplitudes for 200 pixels, averaged over 100 frames, for temperatures between 173-273\,K. In each case, the signal region was fixed (samples 31 to 50 in the 2\,Mpixel/s waveform, see Figure \ref{fig:2MHz_5MHz_WF}), while the baseline region was varied; the last baseline sample was fixed to sample 27, and the first baseline sample was decremented from sample to 26 to sample 11, the last baseline sample before the reset pulse region. The upper horizontal axis of each plot presents the baseline-signal window sampling frequency. The blue curve represents the standard deviations from the waveform obtained using the default bias, while the magenta curve represents the standard deviations from the waveform obtained using the optimized bias at each temperature.

For uncorrelated white noise, it is expected that as the number of samples, $N$, increases, the standard deviation should decrease by a factor of $\sqrt{N}$. However, we see that the standard deviation in the default cases deviates---at some temperatures, dramatically---from this trend. The most drastic examples of this deviation can be seen between 233\,K--263\,K, as the standard deviation begins to increase with more baseline samples beyond a window frequency of around 6.2\,MHz. Interestingly, these temperatures correspond to the resonance peak in the noise (see Fig. \ref{fig:scan_temp_noise_fwhm}). It should be noted that increasing the number of samples in this case also decreases the sampling frequency, and often the lower frequency regime is dominated by $1/f$ noise instead of white noise. This may indicate that the observed increase in the standard deviation of the signal-baseline amplitude for lower sampling frequencies arises from $1/f$ noise. Standard deviations at lower sampling frequencies improve significantly with optimized RG, OG, and RD biases, and correspondingly, the noise resonance strongly depends on these parameters. This implies that the $1/f$ noise is localized in the output stage.

\begin{figure*}[ht!]
   \begin{center}
   \begin{tabular}{c}
   \includegraphics[height=5.5cm]{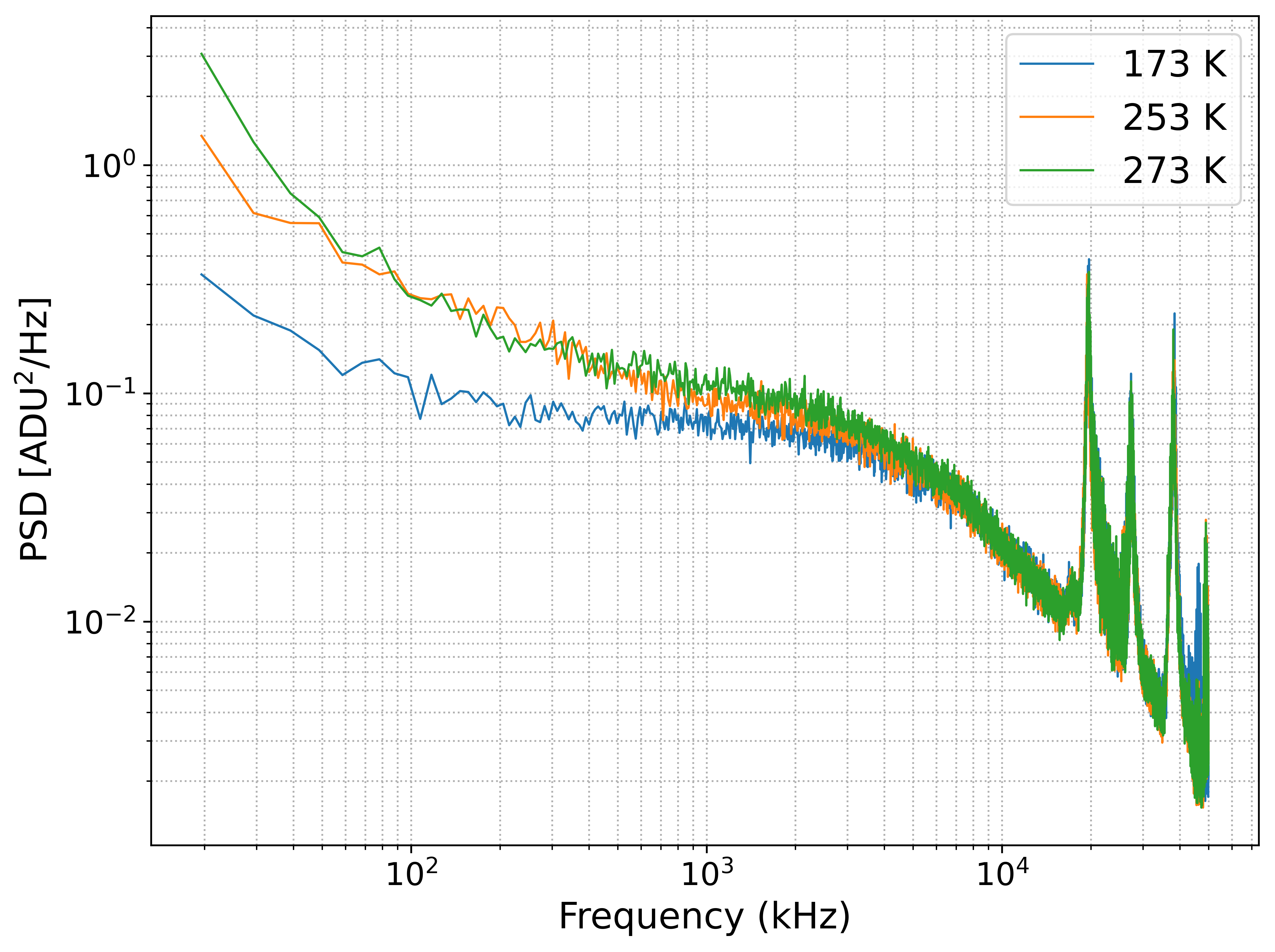}
   \includegraphics[height=5.5cm]{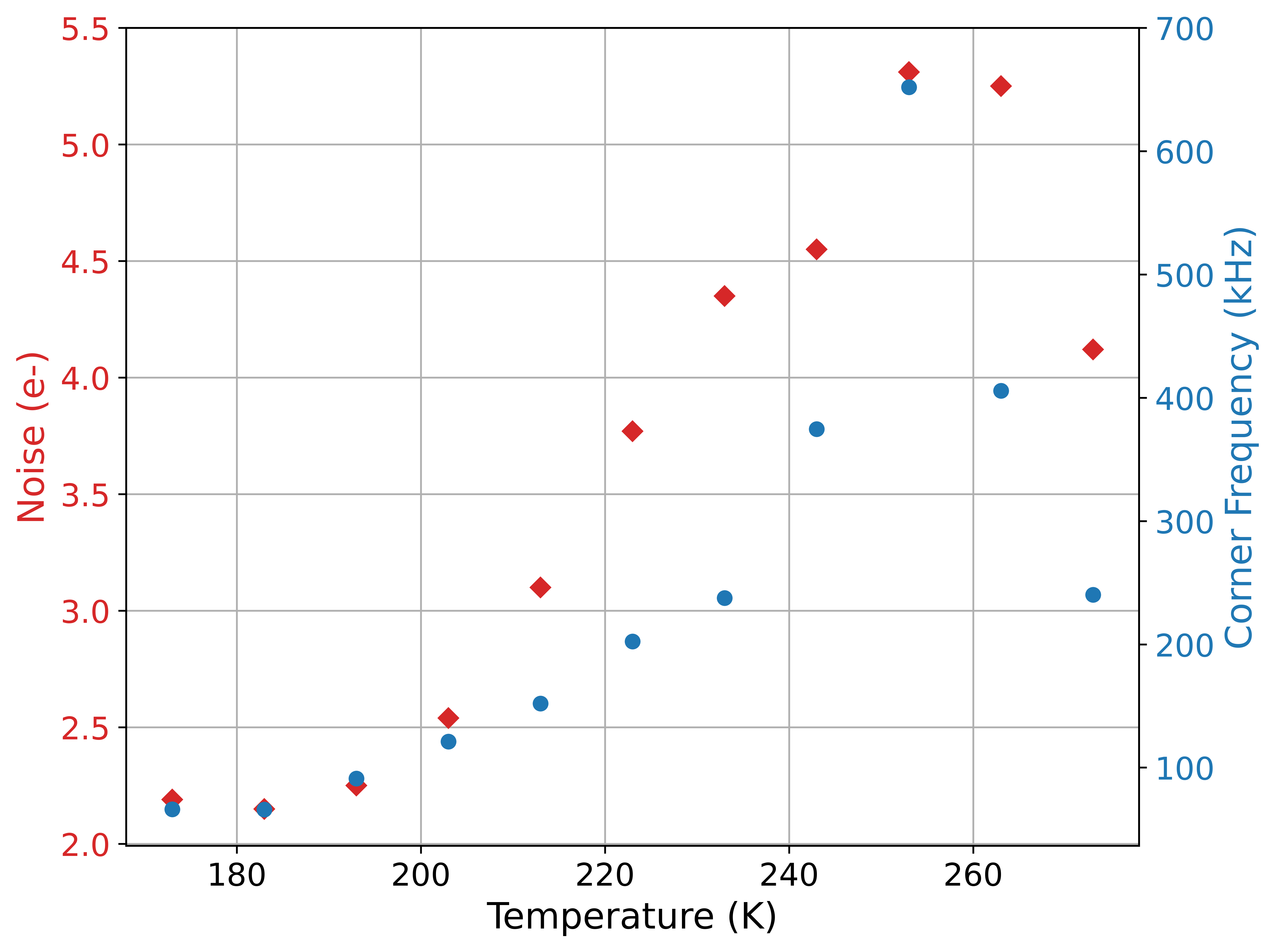}
   \end{tabular}
   \end{center}
   \caption 
   { \label{fig:psd_comparison} 
{\it Left:} Comparison of PSDs obtained at 173\,K (blue), 253\,K (orange), and 273\,K (green), probing the noise behavior of the output stage of the CCID-93 detector.
{\it Right:} Plot of corner frequency, defined as the frequency where the power law component of Equation \ref{eq:psd} is equal to the constant component, and noise in electrons versus temperature. It can be seen that the corner frequency correlates with the noise.
}
\end{figure*}

This localization can be substantiated by examining the PSD of the waveform when the detector is operated in permanent reset, which effectively probes noise from the output stage in isolation. Example PSD plots at 173\,K, 253\,K, and 273\,K can be seen in the left plot of Figure \ref{fig:psd_comparison}. We expect the white noise floor to shift as thermal noise increases with higher temperatures. However, the slope of the noise contribution at lower frequencies also appears to vary between temperatures. 

Using Whittle Likelihood minimization, we fit the 20\,kHz--2000\,kHz range of the PSDs with the model: 
\begin{equation}
\label{eq:psd}
{\rm PSD}(f) = A * f^m + C   
\end{equation}
The model includes a power law component to capture the behavior at low frequencies and a white noise floor for higher frequencies. From these fits, we can determine the intersection frequency of the power law component and the flat component, as a proxy for the contribution of white noise across frequencies. The average fitted power law slope across the datasets was -1.13 with a standard deviation of 0.21. A plot of the crossover frequency obtained from the fitting as a function of temperature is given in the right plot of Figure \ref{fig:psd_comparison}. Also plotted is the read noise in electrons. It can be clearly seen that the electron noise trend is correlated with the crossover frequency, indicating that $1/f$ noise dominates at higher frequencies where the noise is peaking. This picture is consistent with the conclusions drawn from the expanding baseline analysis in Figure \ref{fig:expanding_baseline}. Optimizing the RG and RD bias parameters may alter the output stage channel geometry and charge trap distribution, which in turn changes the trapping timescales, moving the $1/f$ dominated regime of the noise spectrum further from the window sampling frequency. In the next section, we discuss a model of charge trapping in the output stage which demonstrates different noise behaviors with temperature.

\subsection{Output Stage Noise Model}
\label{sec:modeling}

Trapping and detrapping of charge carriers in the output stage transistors can provide a plausible physical origin for the $1/f$ noise observed at the output. Because the trap capture and emission time scales depend strongly on temperature, the contribution of this mechanism to the total noise is expected to vary with both detector temperature and readout speed. In particular, and as we saw when performing the expanding baseline analysis in Section \ref{sec:noise_temp_discussion}, when the characteristic trap time constants become comparable to the CDS sampling time, the resulting random telegraph noise (RTN) can manifest as enhanced low-frequency noise. This phenomenon has been observed in both JFET and MOSFET transistors \cite{Kandiah10.1109, Kandiah1991600, Janesick01}, and may explain the resonant features observed in the noise–temperature scans. To investigate this possibility, we construct a model of the output stage noise that includes both a thermal noise component and a $1/f$ noise component arising from trap dynamics.

\subsubsection{Thermal noise}
The thermal noise of the output stage is assumed to arise primarily from the channel resistance of the p-JFET and the n-MOSFET in the readout chain. The corresponding voltage noise is modeled as
\begin{equation}
e_{\mathrm{thermal}} =
\sqrt{\frac{\gamma~4kT}{g_m}}
\times \sqrt{\mathrm{Bandwidth}} \times \sqrt{2}
\end{equation}
where $T$ is the detector temperature, $k$ is the Boltzmann constant, and $g_m$ is the transconductance of the output stage. The parameter $\gamma$ is the correction factor for the thermal noise density of FET transistors and typically has a value of approximately $2/3$. The $\sqrt{2}$ factor is added to account for the noise from the two amplifier stages, where we assumed the noise from both stages to be identical for the simplicity of the calculations. A bandwidth of 60\,MHz is adopted for the calculation of the total noise, corresponding approximately to the experimentally measured bandwidth of the combined output stage and readout electronics.
The thermal noise is treated as a Gaussian process whose variance scales with $1/g_m$ and detector temperature. Using the known electronic conversion factor, the noise is converted from voltage to ADU. The resulting thermal noise sequence is then added to a simulated detector waveform corresponding to the chosen readout speed (1--4\,MPixel/s). The waveform is represented as a step pulse, where the high state corresponds to the baseline and the low state corresponds to the signal level, sampled with a 10\,ns time resolution consistent with the 100\,MHz analog-to-digital converter used in the readout electronics.

\subsubsection{Trap-induced 1/f noise}
The $1/f$ noise component is modeled as an RTN process arising from trapping and detrapping of carriers. For a single trap, the RTN signal is represented as a time series that switches between a trapped state (signal low state, state = 0) and a detrapped state (signal high state, state = 1) with a time resolution of 10\,ns.
The state transitions are generated using a Monte Carlo method. At each time step, the probability of trapping or detrapping is computed from the characteristic trap time constant which is modeled as
\begin{equation}
\tau =
\frac{1}{\sigma v_{\mathrm{th}} N_c}
~e^{\left(\frac{\Delta E}{kT}\right)}
\end{equation}
where $\sigma$ is the trap capture cross section (cm$^{2}$), $v_{\mathrm{th}}$ is the electron thermal velocity, $N_c$ is the effective density of states in the conduction band ($2.8\times10^{19}$ cm$^{-3}$), and $\Delta E = E_{\mathrm{cond}} - E_{\mathrm{trap}}$ is the trap energy level relative to the conduction band.
The probability of a state transition, assumed for simplicity to be the same whether it be a trapping or detrapping event, within a time step $dt$ is then given by
\begin{equation}
p = 1 - e^{-dt/\tau}
\end{equation}
where $dt$ is the simulation time step (10\,ns). There is a fixed number of states in the simulation, and the number of occupied states is retained in memory between steps. At each step, we draw a number from a uniform distribution between 0 and 1. If the probability $p$ of a transition is greater than that number, a state transition occurs.
For the RTN model, traps are assumed to occupy a single energy level. To simulate $1/f$ noise, however, a distribution of traps with energies uniformly distributed between two energy bounds is assumed. 
In that case, for each trap, the energy level is chosen randomly between the two energy bounds and used to calculate the value of the time constant. 
The RTN contribution from each trap is generated independently, and the contributions from all the traps ($N_\mathrm{trap}$) are summed to produce the final $1/f$ noise signal.
The resulting noise signal is converted to ADU using the known electronic conversion factor and added to the modeled detector waveform.

\subsubsection{Noise extraction}
\begin{figure}
    \centering
    \includegraphics[width=0.49\textwidth]{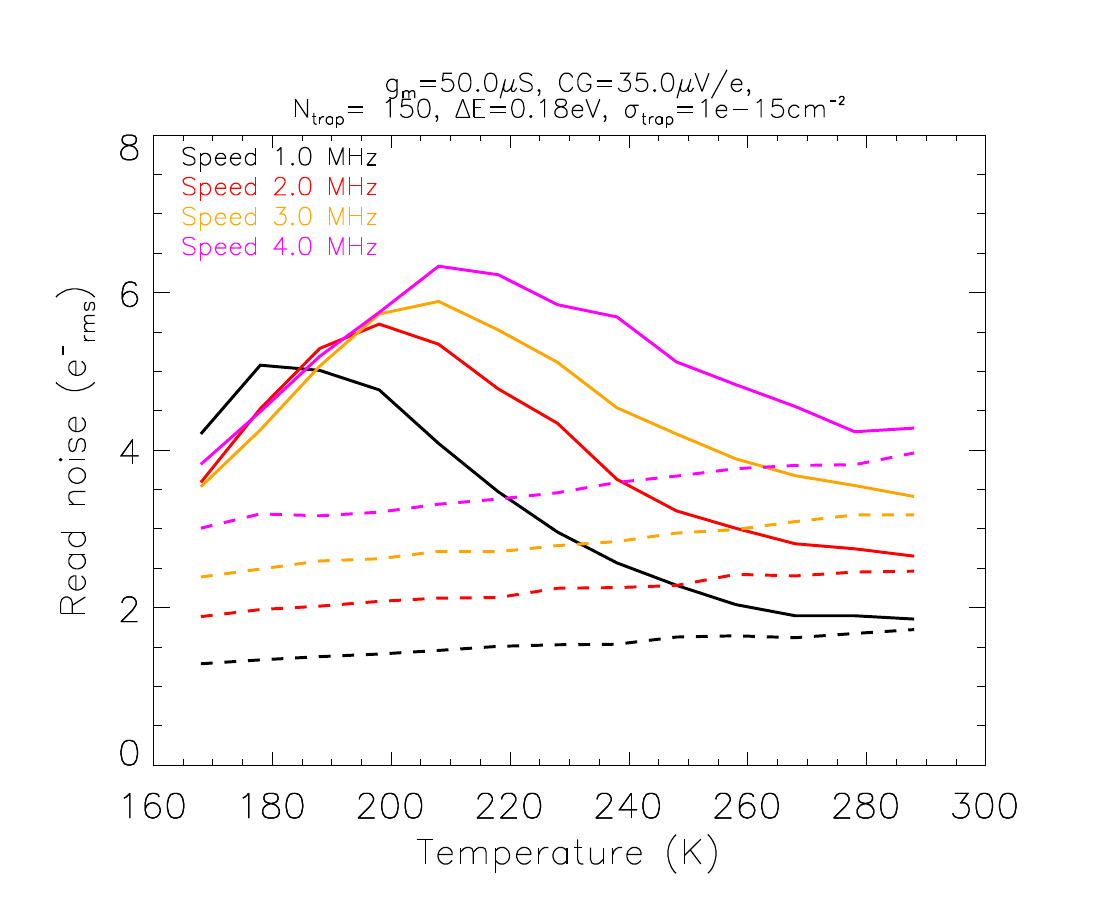}
    \includegraphics[width=0.49\textwidth]{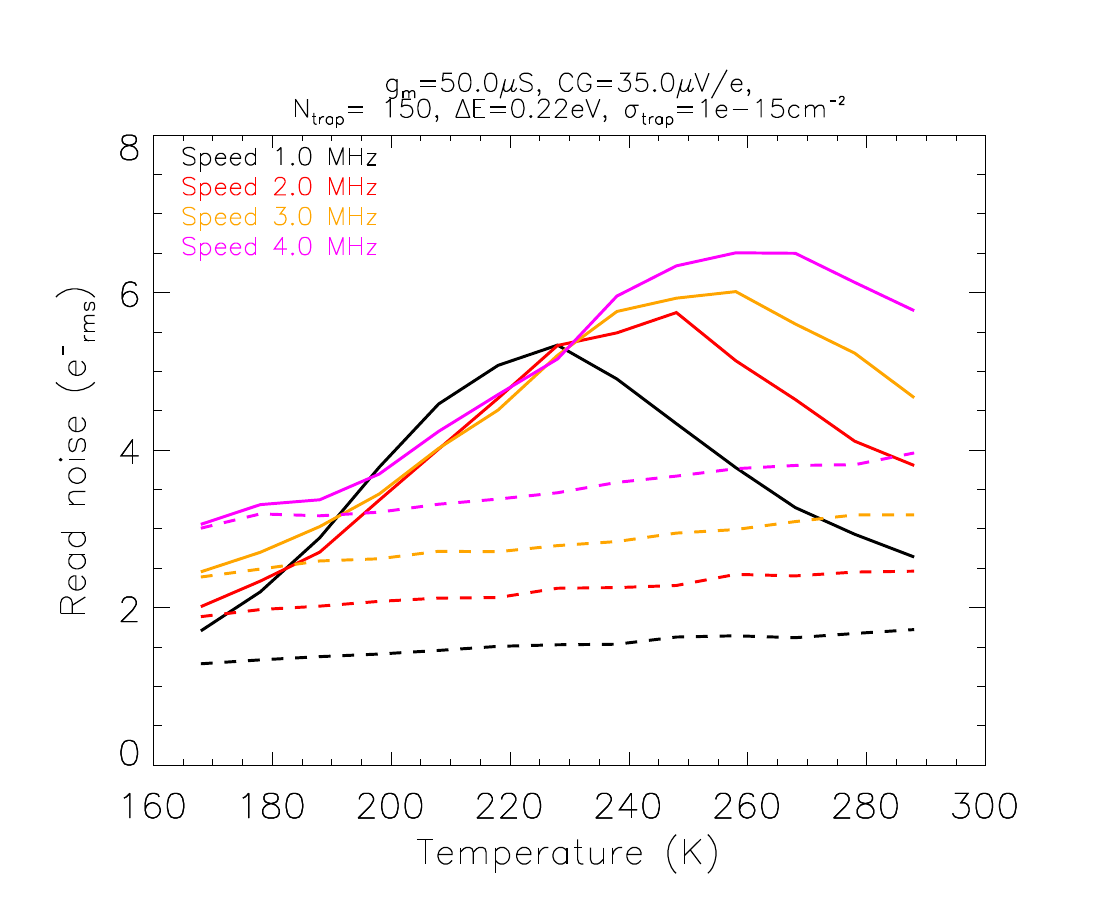}\\
     \includegraphics[width=0.49\textwidth]{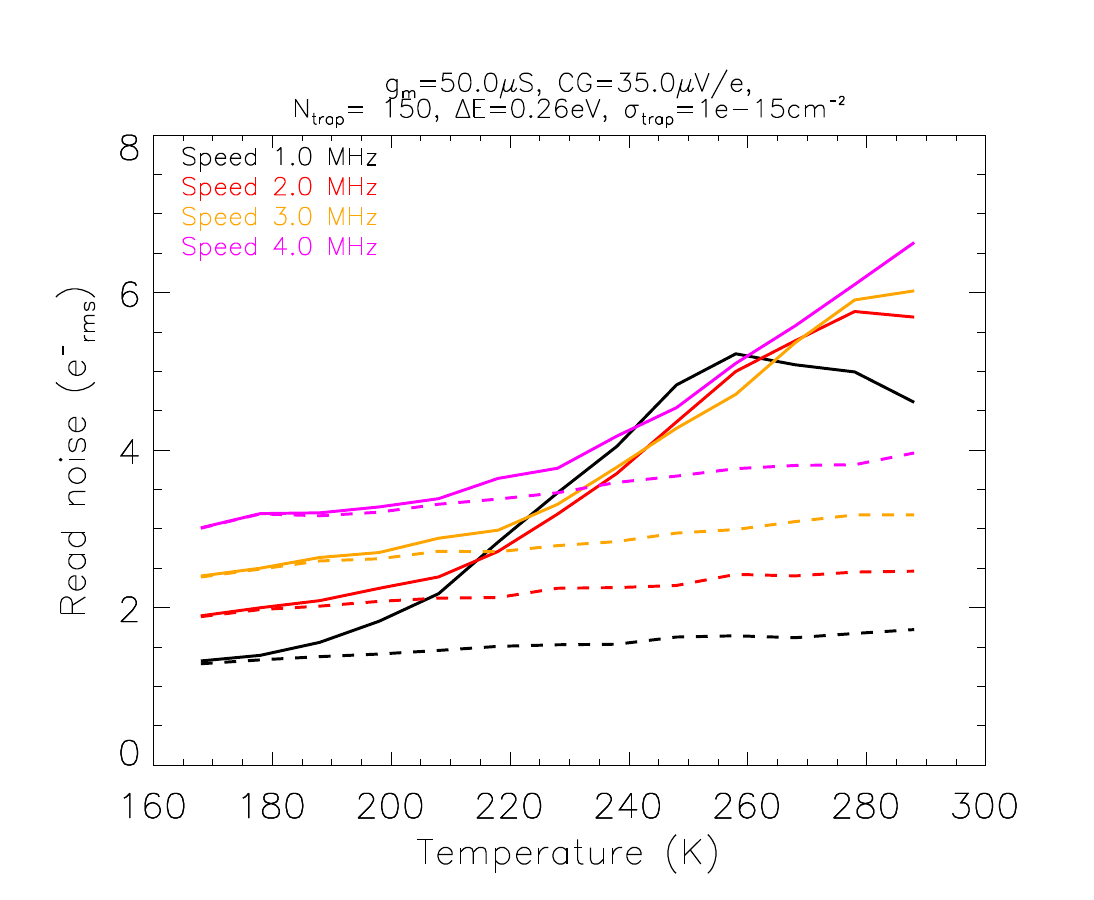}
    \includegraphics[width=0.49\textwidth]{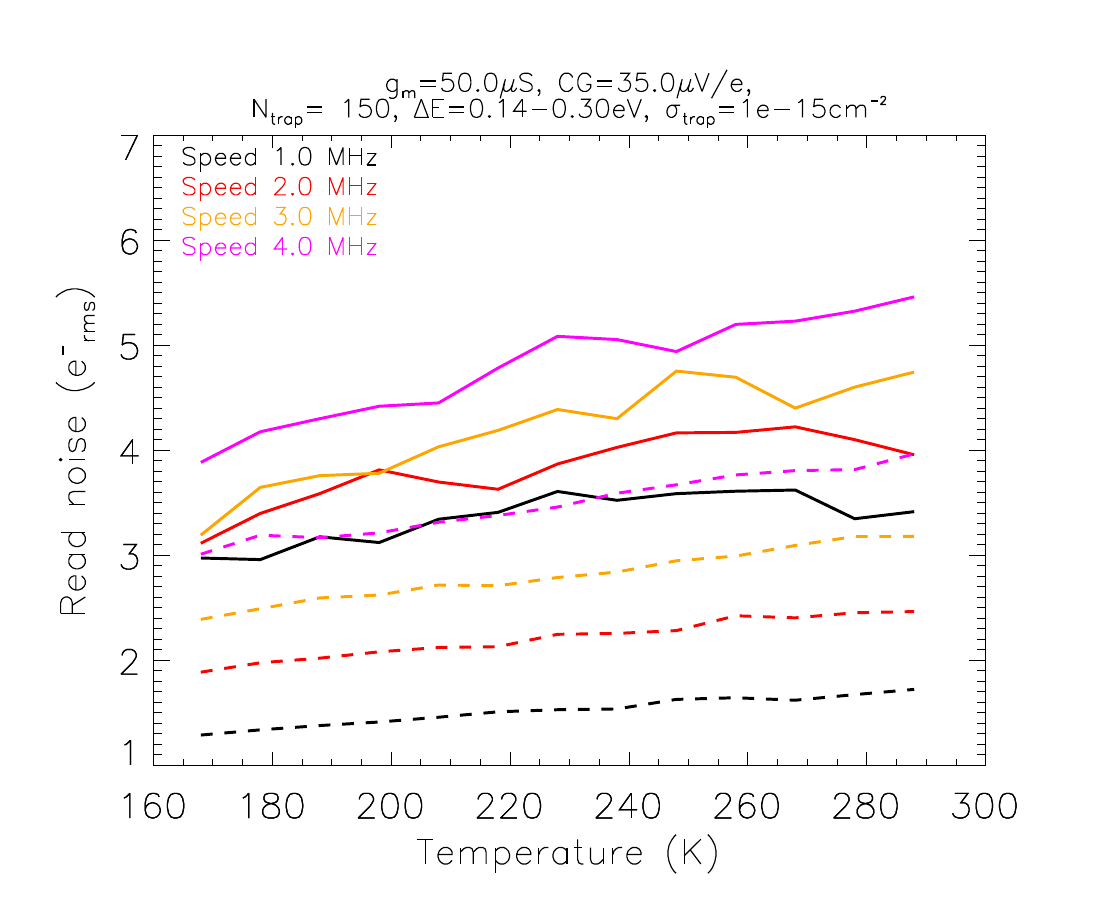}
    \caption{Simulated noise from the output-stage noise model. Top left: thermal (dashed lines) and total (solid lines) read noise as a function of temperature and readout speed of the detector for $\sigma=$~10$^{-15}~\mathrm{cm^2}$, $\Delta E=$~0.18 eV, and $N_\mathrm{trap}=$~150 ($3.75\times10^{9}~\mathrm{cm^{-2}}$ trap density). Top right: same as top left but for $\Delta E=$~0.22\,eV. Bottom left: same as top left but for $\Delta E=$~0.26\,eV. Bottom right: here the traps are assumed to be uniformly distributed between 0.14 and 0.3\,eV relative to the conduction band ($2.34\times10^{10}~\mathrm{cm^{-2}\,eV^{-1}}$).}
    \label{fig:trap_model_noise}
\end{figure}

The simulated waveform is processed using the same procedure applied to the experimental data. The baseline and signal regions are sampled and a CDS operation is performed to compute the pixel signal. The read noise is then estimated as the standard deviation of the resulting pixel signal values. The read noise value is finally converted to electrons using the known detector conversion gain (CG). Figure \ref{fig:trap_model_noise} shows the simulated read noise as a function of temperature and detector readout speeds assuming $N_\mathrm{trap} = 150$ and for three different trap energy levels ($\Delta E=$~0.18 eV, 0.22 eV, 0.26 eV) relative to the conduction band and for a trap cross-section of 10$^{-15}~\mathrm{cm^2}$. Assuming interface traps are the primary contributor to the 1/f noise component, and with a gate size of $\sim2\,\rm \mu m \times 2\,\rm \mu m$, this would correspond to a trap density of $3.75\times10^{9}~\mathrm{cm^{-2}}$.

\begin{figure}
    \centering
    \includegraphics[width=0.49\textwidth]{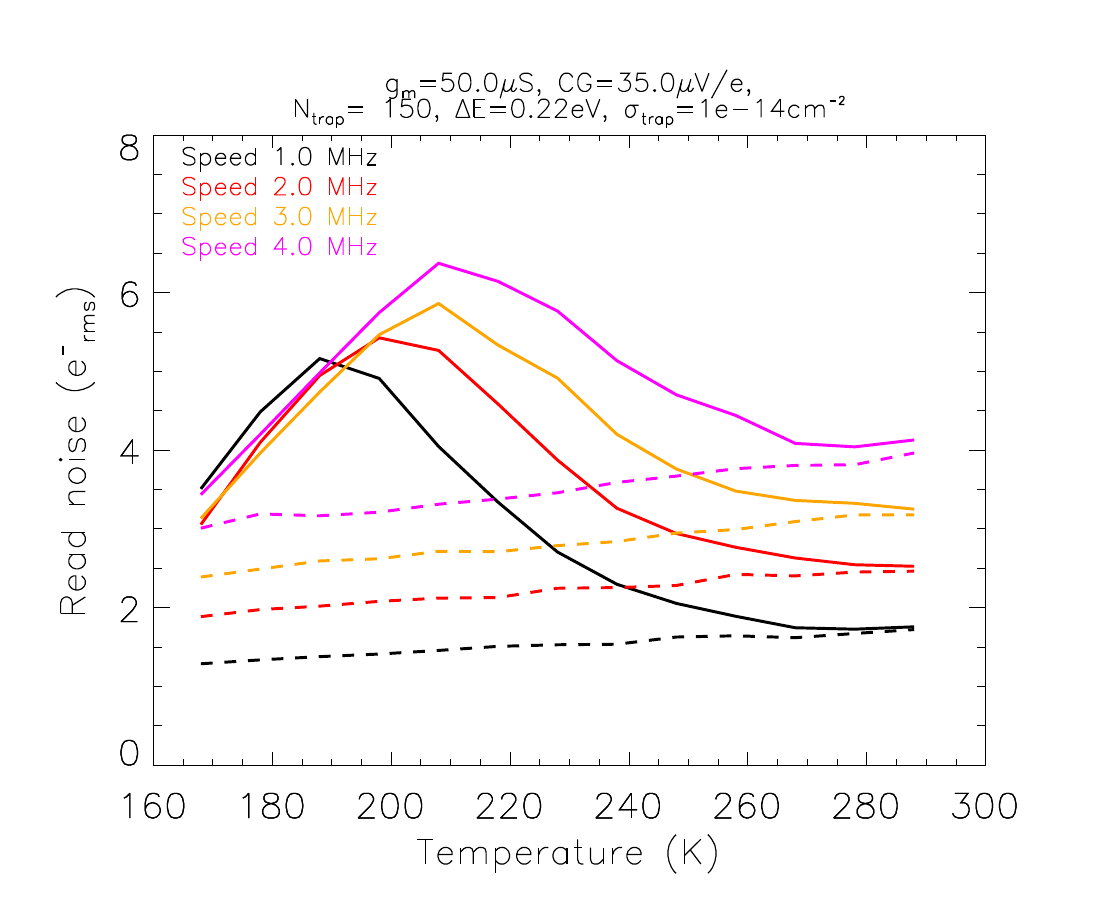}
    \includegraphics[width=0.49\textwidth]{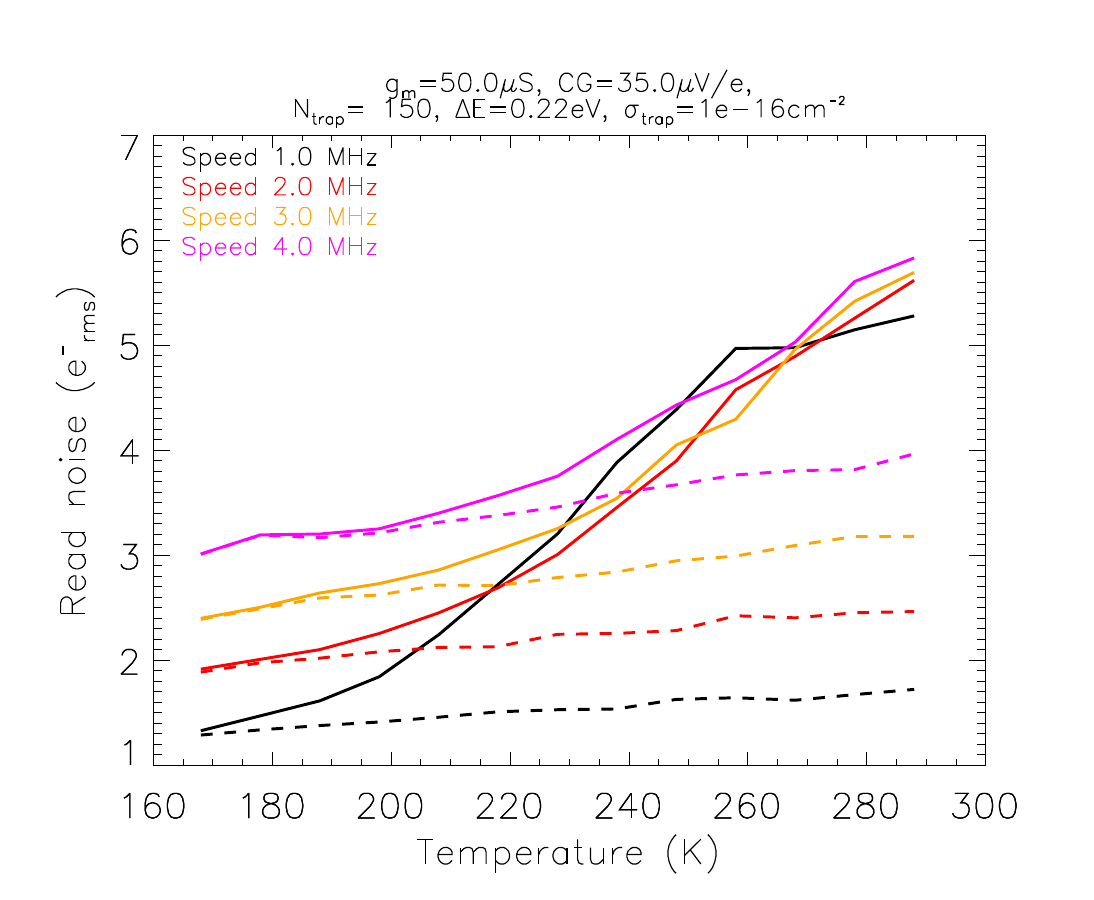}
    \caption{Same as Fig. \ref{fig:trap_model_noise} but for two different trap cross-sections: $\sigma=$~10$^{-14}~\mathrm{cm^2}$ (left) and $\sigma=$~10$^{-16}~\mathrm{cm^2}$ (right) for $\Delta E=$~0.22\,eV.}
    \label{fig:trap_model_cross}
\end{figure}

As shown by the solid lines in these figures, the model reproduces the resonant features in the total noise at certain temperatures that are present in the real data in Figure \ref{fig:scan_temp_noise_fwhm}. The dashed lines represent the thermal noise component in the total noise. The resonance peak moves to higher temperatures for higher readout speeds, which is expected because at higher temperatures the trap/detrap time scale is shorter, making the resonance possible at shorter sampling times of the waveform. The resonance features are found to smear out significantly if we consider the traps to be uniformly distributed in a broader energy range ($1/f$ noise) as shown in the bottom right panel of Fig. \ref{fig:trap_model_noise}. Figure \ref{fig:trap_model_cross} shows the effect of weaker or stronger trap cross-sections on the noise resonance features. 

These results suggest that the traps in our CCD output stage are shallow ($<$0.3\,eV relative to the conduction band), localized in energy space and weakly interacting (cross-section in the range 10$^{-14}~\mathrm{cm^2}$ to 10$^{-16}~\mathrm{cm^2}$), i.e. neutral or weakly charged. Vacancy-oxygen complex or VO traps, which are introduced when a silicon vacancy (due to irradiation or ion implantation) is occupied by interstitial oxygen, are known to be neutral, have a low cross-section, and typically have energies close to the conduction band \cite{furuhashi2005_trap}. High concentrations of oxygen during wafer growth can form a VO complex. Further investigation on the cause of traps during device fabrication and its mitigation is in progress and being investigated for other detector formats using the same output stage architecture.

\section{Summary}
\label{sec:conclusion}

The XOC group at Stanford University with collaborators at MIT-LL and MKI are working towards enabling the frame rates and noise performance requirements of next-generation X-ray imaging observatories. The combination of the MIT-LL single-poly CCID-93 CCD with our dedicated MCRC V1.0 readout ASIC, and the integration of fast clock drivers (specifically for SW and RG) close to the CCD, enables order of magnitude faster readout speeds than legacy X-ray CCDs. 

We have developed an automatized optimization process for CCD bias voltages (RG, OG and RD) that finds the best operating points (and therefore noise performance) efficiently. The combination of these approaches allows us to reliably deliver high frame rates with excellent readout noise. Furthermore, we have presented raw waveform and PSD analysis techniques for diagnosing and characterizing residual noise sources in these detectors, presenting a physical model for $1/f$ noise sources in the two-stage output of the CCID-93 MIT-LL detectors.

Our results demonstrate clearly that the MIT-LL CCD fabrication process and state-of-the-art electronics can deliver the speed and noise performance required of future strategic X-ray missions. We are now focusing efforts to characterize and optimize larger-format, 16-channel 1440$\times$1440 pixel CCID-100 devices, which have the same output stage configuration as the CCID-93. This process will be streamlined by the methods presented here, and similar procedures should be applicable to any future strategic X-ray mission requiring fast, low-noise, large-format imaging detectors.

\section{Disclosures}
The authors declare that there are no financial interests, commercial affiliations, or other potential conflicts of interest that could have influenced the objectivity of this research or the writing of this paper. This work expands upon an SPIE Optics + Photonics 2025 conference manuscript.\cite{StueberSPIE2025}

\section{Acknowledgments}

We acknowledge support from NASA APRA grant 80NSSC22K1921 and NASA SAT grant 
\newline 80NNSC23K0211, as well as support from NASA for the AXIS Probe Phase A study, under contract 80GSFC25CA019. ChatGPT (OpenAI) was used for non-substantive coding assistance. It was not used to perform analysis or to generate figures or scientific content. The authors reviewed and verified all code, and remain fully responsible for the final content.
      
\clearpage
% References
% \bibliography{report} % bibliography data in report.bib

\bibliographystyle{spiebib} % makes bibtex use spiebib.bst

\end{spacing}
\end{document}